# Decoding the RVb Enigma of DF Cyg: Pulsations, Circumstellar Disks, and Post-RGB Evolution Revealed by Multi-Wavelength Observations

**Cenk Kayhan**

Scientific Research Projects Coordination Unit, Kayseri University, Kayseri 38280, Türkiye

Department of Astronomy and Space Sciences, Faculty of Science, Erciyes University, Kayseri 38030, Türkiye

E-mail: `cenkkayhan@kayseri.edu.tr`

**Timur Şahin**

Department of Space Sciences and Technologies, Faculty of Science, Blok-B, Akdeniz University, Antalya 07058, Türkiye

**Abstract.** DF Cyg, an extensively studied RV Tauri star characterized by irregular pulsation dynamics, represents the sole RV Tauri star within the original Kepler field. This investigation integrates five decades of ground-based photometric data from the *AAVSO, AFOEV, ASAS, ASAS-SN*, and *SuperWASP* surveys with recently acquired *TESS* observations, conducting a multi-wavelength and multi-phase photometric analysis to elucidate the complex nature of DF Cyg. For the first time, we analyze approximately four years of *Kepler* space telescope data alongside *TESS* observations from sectors 14–15, 41, and 54–55. The precision of the *TESS* Sector 14 data approaches that of the *Kepler* dataset. Furthermore, *TESS* observations from sectors 15, 54, and 55, obtained in 120-second cadence mode, constitute the highest-precision photometric dataset for DF Cyg to date. By isolating long-term trends in the *TESS* data, we quantified short-term variations in the fundamental pulsation frequency and its integer harmonics. The two alternating short-term cycles in the *TESS* light curve facilitated the unambiguous identification of seven previously undetected integer harmonics of the fundamental frequency in the power spectrum (2f/9, 3f, 4f, 5f, 6f, 7f, and 8f), providing critical new insights into the star's complex pulsation dynamics. A periodogram analysis of the combined ground- and space-based datasets revealed approximately 35 frequencies linked to both long- and short-term variability mechanisms. Stellar parameters derived from Gaia DR3 data—specifically, luminosity and radius estimates based on Type II Cepheid period-luminosity (PL) and period-radius (PR) relations—demonstrate consistency with values obtained through spectral energy distribution (SED) modeling. Complementary high-resolution spectroscopic analysis of DF Cyg, conducted using the SP_Ace code on McDonald Observatory data, yielded atmospheric parameters of $T_{\rm eff} = 4781$ K, $\log g = 1.74$ cm s$^{-2}$, and $[Fe/H] = -0.07$ dex, further constraining the star's physical characteristics. This synthesis of multi-epoch, multi-instrument data advances the understanding of pulsational behavior and evolutionary status in RV Tauri systems. Our findings highlight DF Cyg as a critical benchmark for its class, as its characteristics bridge the gap between post-RGB systems in the Magellanic Clouds and higher-luminosity post-AGB stars.

# 1. Introduction

As members of the Type II Cepheid class, RV Tauri stars occupy the high-luminosity regime of the Hertzsprung-Russell (HR) diagram (Wallerstein, 2002). These stars exhibit large-amplitude photometric variations, typically characterized by alternating deep and shallow minima in their light curves. Photometric studies estimate the interval between successive deep minima to range from 30 to 150 days (e.g., Pollard et al., 1997; Percy, 2007). Notably, Soszyński et al. (2008) identified the shortest pulsation period (21.4829509 d) for an RV Tauri star, T2CEP-067 in the Large Magellanic Cloud, which arises from radial pulsations associated with post-red giant branch (RGB) or asymptotic giant branch (AGB) evolutionary phases (Jura, 1986).

Long-term brightness variations, spanning 700–2500 d, further categorize RV Tauri stars into two subgroups: RVa (constant mean brightness) and RVb (variable mean brightness). These RVb variations are attributed to mechanisms such as binary occultations (Van Winckel et al., 1999), circumstellar dust interactions (Kiss and Bódi, 2017), or intrinsic stellar processes (Zsoldos, 1996; Lloyd Evans, 1985). Despite their significance, the RVb phenomenon remains inadequately explained. DF Cyg, a well-documented RVb star, has been extensively studied (Belserene, 1984; Vega, Montez, Stassun and Boyd, 2017; Bódi et al., 2016; Vega, Stassun, Montez, Boyd and Somers, 2017; Plachy et al., 2018; Manick et al., 2019) and is uniquely positioned as the sole RV Tauri star within the original *Kepler* field (Bódi et al., 2016; Borucki et al., 2010).

DF Cyg was first identified by Harwood (1927), who reported a magnitude range of $10^m.7$–$14^m.2$ and a period of 49.4 days. Townley et al. (1928) classified it as an RV Tauri variable, noting irregularities in its light curve. Subsequent photometric analyses by Belserene (1984), Howard (1989), and Percy and Ursprung (2006) revealed complex periodicity variations, including short-term ($\approx$49.8 days) and long-term ($\approx$775 days) cycles.

High-precision space photometry has significantly advanced DF Cyg research. Bódi et al. (2016) combined 4 years of *Kepler* data with 50 years of AAVSO observations, confirming short- and long-term periods of 49.85 and 779.606 days, respectively. Their work highlighted nonlinear pulsations and underscored the need for theoretical modeling (Borucki et al., 2010). Similarly, Vega, Stassun, Montez, Boyd and Somers (2017) interpreted the long-term period (795 ± 5 days) as evidence of binarity, attributing flux variations to disk occultation by a companion. This interpretation aligns with SED features indicative of a circumbinary disk (Vega, Stassun, Montez, Boyd and Somers, 2017) and is supported by Kluska et al. (2022a), who associate Galactic RVb stars with high-inclination systems.

Vega, Stassun, Montez, Boyd and Somers (2017) derived a radius of $10.3 \pm 3.8$ R$_\odot$ and a distance of $990 \pm 372$ pc using Gaia parallax data, alongside a pulsation period of $49.84 \pm 0.02$ days from *Kepler* observations. Their methodology included correcting discontinuities in the *Kepler* long-cadence SAP data‡ by adjusting quarters 0–3, 9, and 12. Spectroscopic and photometric studies by Manick et al. (2019) further proposed DF Cyg as a post-RGB binary system (a low-luminosity analog of post-AGB stars) with a $0.6 \pm 0.1$ M$_\odot$ companion,

---

‡ https://archive.stsci.edu/kepler/

emphasizing its infrared excess and complex atmospheric properties.

Kluska et al. (2022b) categorized DF Cyg as a "Category 0" system with a highly inclined disk, based on infrared color indices (W1–W3 and H–$K_s$). Kluska et al. (2022b) classified 85 Galactic post-AGB stars into five categories (0–4) based on their positions within the *(W1–W3)* versus *(H–Ks)* color–color diagram. Category 0, defined by the criteria $H-Ks > 1.15$ and $W1-W3 < 4.5$, comprises targets exhibiting the highest infrared-to-stellar luminosity ratios ($> 100\%$). This group occupies the upper region of the diagram, characterized by elevated near-infrared colors ($H-Ks > 1.15$). Furthermore, Category 0 contains the largest cohort of targets exhibiting the RVb phenomenon among the classified categories.

Recent work by Mohorian et al. (2024) investigated the star's photospheric composition using HERMES/Mercator and APOGEE spectra, suggesting binary-driven evolutionary alterations.

This study presents the first analysis of DF Cyg using *TESS* data, complementing 50 years of ground-based photometry (AAVSO, AFOEV, ASAS, ASAS-SN, SuperWASP) and space-based observations (*Kepler*, *TESS*) to conduct a multi-wavelength, multi-phase investigation of its long- and short-term variability. We aim to elucidate the mechanisms underlying the RVb phenomenon through this comprehensive dataset. The paper is structured as follows: Section 2 details observational data sources, Section 3 outlines photometric and spectroscopic methodologies, Section 4 applies period-luminosity and period-radius relations, Section 5 discusses implications, and Section 6 summarizes conclusions.

## 2. Data

### 2.1. Photometric Observations

*2.1.1. Kepler Space Telescope:* DF Cyg (KIC 7466053) was observed continuously by the *Kepler* space telescope (Borucki et al., 2010) across 18 quarters. For this study, *Kepler* Target Pixel Files (TPFs) for each quarter were retrieved from the Mikulski Archive for Space Telescopes (MAST) and processed into light curves. The dataset spans nearly four years (Q0–Q17), with the target's point spread function (PSF) distributing its flux across multiple CCD pixels. To optimize flux extraction, custom aperture masks were constructed for each quarter to encapsulate all stellar flux while minimizing contamination from background sources and adjacent pixels. Instrumental artifacts and noise were systematically mitigated using methodologies detailed by Hartig et al. (2014).

Owing to variations in CCD channel positioning across quarters, individual light curves exhibited distinct flux baselines. These were combined into a unified time series for long-term analysis, with outliers excised during post-processing. The final composite light curve spans approximately 1450 days, featuring gaps of $\sim$1.5 days due to *Kepler*'s solar alignment maneuvers. All observations were conducted in long-cadence mode (30-minute integration). Figure 1 presents the full *Kepler* light curve, binned into three segments (Q0–2/Q9–11, Q3–5/Q12–14, Q6–8/Q15–17) to illustrate long-term brightness evolution. The first panel demonstrates a gradual rise to maximum brightness, the second panel shows a plateau phase, and the third panel depicts a decline to minimum brightness, quantified in *Kepler* magnitudes ($K_p$).

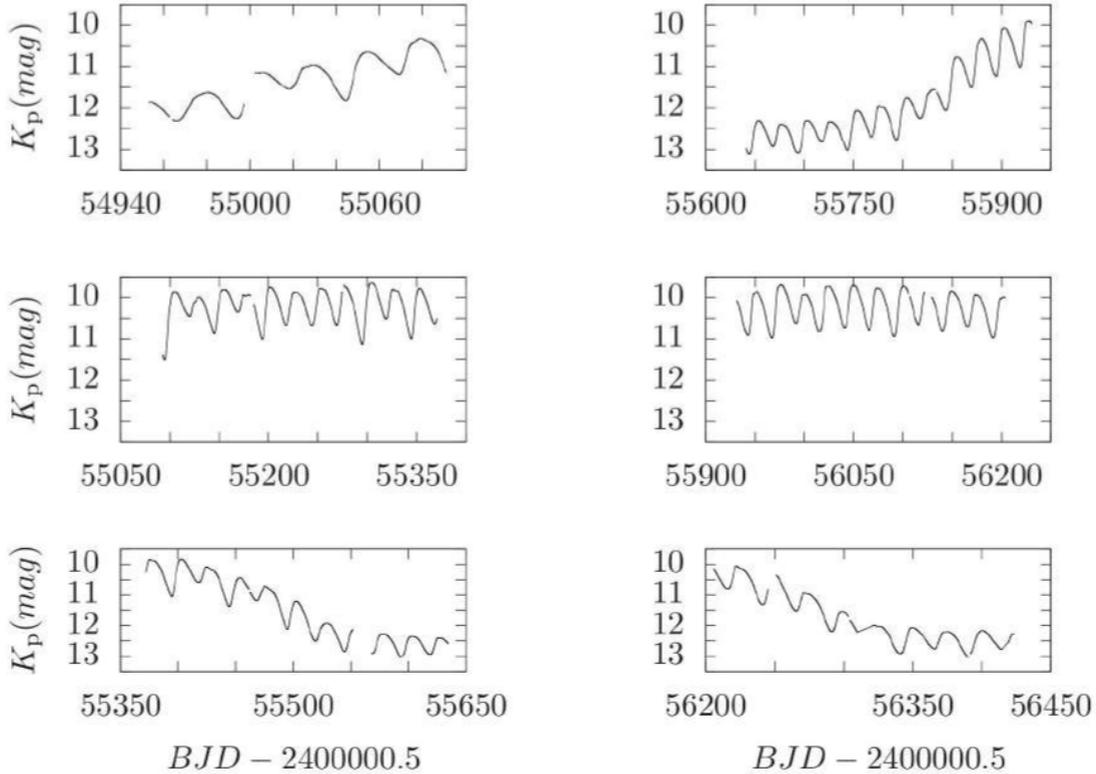

**Figure 1.** Derived light curve from all quarters of *Kepler* data. Each plot included three-quarters of the data in the sub-plots. In the first row of the multiplot (Q0-2 and Q9-11), the magnitude of DF Cyg is increased to the maximum brightness level. DF Cyg is reached its maximum brightness level and has nearly at an almost constant brightness value in the aspect of long-term variations in the second row of the multiplot (Q3-5 and Q12-14). Lastly, in the third row of the multiplot (Q6-8 and Q15-17), the brightness of DF Cyg is decreased to the lowest level.

Methodological divergence from prior studies is noteworthy. While Bódi et al. (2016) sourced raw light curves from the *Kepler* Asteroseismic Science Operations Center§ (KASOC) and applied techniques akin to Bányai et al. (2013), this work derives light curves directly from TPFs using the Hartig et al. (2014) pipeline, resulting in discrepancies illustrated in Figures 7 (Q0) and 8 (Q1–Q17). Although Vega, Stassun, Montez, Boyd and Somers (2017) utilized identical Simple Aperture Photometry (SAP) data, they corrected systematic discontinuities in quarters Q0–3, Q9, and Q12. Such adjustments were intentionally omitted here to preserve the integrity of both long- and short-term variability signals for unbiased analysis.

*2.1.2. TESS:* To investigate the short-term variability of DF Cyg (TIC 272951532), we employed high-precision photometric data from the *Transiting Exoplanet Survey Satellite*

§ https://kasoc.phys.au.dk/

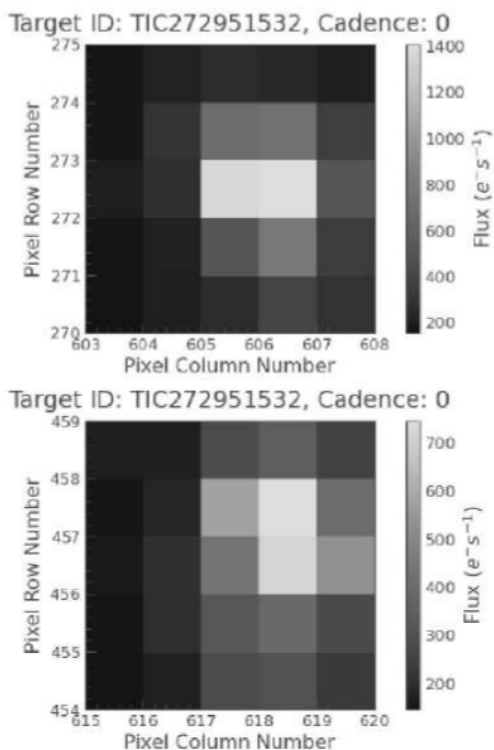

**Figure 2.** The *TESS* Target Pixel Files (TPFs) of DF Cyg (TIC 272951532) in sectors 14 (top) and 15 (bottom). Bright pixels of the target were used to obtain light curves. The dark pixels associated with the background are removed. The same procedure is applied to sectors 54 and 55.

(*TESS*) (Ricker et al., 2015). The star was observed in five sectors (14, 15, 41, 54, and 55), with sectors 14–15, 41, and 54–55 analyzed for the first time in this study. Due to its position near the edge of the full-frame image (FFI) in Sector 41, these data were excluded from analysis. High-cadence observations (120-second mode) were available exclusively in Sectors 15, 54, and 55.

A data reduction methodology analogous to that outlined in Section 2.1.1 was applied to the *TESS* data. Target Pixel Files (TPFs) for each sector, retrieved from the MAST database, were processed into light curves using custom aperture masks and background subtraction. Background flux was systematically subtracted from individual pixels to minimize noise contributions. Instrumental artifacts were further mitigated using the LIGHTKURVE software package (Lightkurve Collaboration et al., 2018).

Figure 2 displays the light curves derived from Sectors 14 and 15. The combined dataset from all four usable sectors comprises over 25,000 photometric measurements. However, the discontinuous and limited temporal coverage of individual *TESS* sectors (∼27 days) restricts analysis to short-term variability. Long-term trends, such as those identified in *Kepler* data, are not resolvable within the *TESS* observational baseline.

**Table 1.** Summary of ground-based photometric surveys contributing to the analysis of DF Cyg, including filters, observational time spans, number of data points, and primary contributions to short- and long-term variability studies.

| Telescope/Survey | Filter(s) | Time Span | Number of Data Points | Key Contribution |
|---|---|---|---|---|
| ASAS | V, I | July 2006 - Dec 2007 | ∼200 | Dense coverage for short-term variations |
| SuperWASP | 400-700 nm | June 2006 - Aug 2008 | ∼4400 | Large number of data points for detailed analysis |
| ASAS-SN | V | Feb 2015 - Nov 2017 | ∼70 | Longer time span for studying long-term trends |

*2.1.3. Ground-base Observations:* In addition to space-based photometry, ground-based observations were incorporated to achieve a holistic characterization of DF Cyg's variability. The American Association of Variable Star Observers (AAVSO) (Kafka, 2020) has compiled over 7430 observations of DF Cyg since 1968, spanning nearly five decades. These data include visual magnitudes (precision: ±0.5 mag) and measurements in Johnson B, V, and Cousins I filters. While the AAVSO dataset ‖ exhibits temporal discontinuities exceeding the star's fundamental period, long-term variability trends were extracted via data binning at intervals below half the short-period cycle.

The French Association of Variable Star Observers (AFOEV) (Schweitzer and Vialle, 1993) complements this record with 5316 data points spanning 1927–2020¶. Despite temporal gaps (1932–1971; 1973–1976), the AFOEV dataset provides historical context through visual observations and limited photometric data in Johnson B (9 CCD), V (224 photoelectric, 443 CCD), and Cousins R (1584 photoelectric, 7 CCD) filters. Identical processing methods to the AAVSO data were applied to ensure consistency in period analysis.

The All Sky Automated Survey (ASAS) monitored DF Cyg (ASAS_ID 194854+4302.2) in Johnson V and Cousins I bands from July 2006 to December 2007, yielding ∼90 and ∼110 data points, respectively. This dense temporal sampling facilitates short-term variability studies. Similarly, the SuperWASP survey acquired nearly 4400 observations between June 2006 and August 2008, predominantly in 2008, enabling high-cadence analysis of pulsational behavior.

The All-Sky Automated Survey for Supernovae (ASAS-SN) contributed observations of DF Cyg (ASASSN-V J194853.94+430214.5) from February 2015 to November 2017 (Pawlak et al., 2019). This extended temporal baseline supports investigations into long-term brightness trends, complementing the high-cadence ground- and space-based datasets.

*2.2. Spectroscopic Observations*

To complement the photometric analysis, high-resolution spectroscopic observations of DF Cyg were obtained using the 2.1 m Struve reflector telescope at W. J. McDonald Observatory, equipped with the Sandiford Cassegrain echelle spectrograph (McCarthy et al., 1993). Data were acquired over four nights (16 and 18 September; 16 November; 19 October) in 2008, covering the wavelength range 4800–5600 Å to probe elemental abundances and atmospheric properties. A spectral resolution of ∼55 000 was achieved.

---

‖ https://app.aavso.org/webobs/results/?star=000-BCJ-079
¶ http://cdsarc.u-strasbg.fr/cgi-bin/afoevList?cyg/df

Standard reduction procedures were implemented using the Image Reduction and Analysis Facility (IRAF) (Tody, 1986, 1993). These included bias subtraction via polynomial modeling of the overscan region, flat-field correction using halogen lamp exposures to address pixel sensitivity variations, scattered light removal, cosmic ray rejection, and wavelength calibration with Th-Ar arc spectra obtained before and after each stellar exposure. Data quality was monitored through daytime sky observations and radial velocity (RV) standard star measurements. The wavelength calibration achieved an internal precision of 0.002 Å, corresponding to an RV accuracy of $\sim$109 m s$^{-1}$. Signal-to-noise ratios (SNR) varied across echelle orders, peaking at $\sim$20. Individual orders were normalized and merged into a continuous spectrum using the LIME code (Şahin, 2017) within the *Interactive Data Language (IDL)*[+].

While these spectra offer critical constraints on atmospheric parameters and chemical composition, the present study prioritizes photometric analysis, with spectroscopic results serving as ancillary diagnostics.

## 3. Analysis of Photometric and Spectroscopic Data

### 3.1. Analysis of Light Curves

To characterize the short- and long-term variability of DF Cyg, we applied the Lomb-Scargle periodogram[*] (Lomb, 1976; Scargle, 1982). Given the RVb-type variability of DF Cyg, individual light curves were analyzed to isolate pulsation frequencies linked to short-term dynamics. Frequency peaks with a signal-to-noise ratio (SNR) exceeding 4.0 were deemed statistically significant, consistent with established thresholds for time-domain astronomy. Amplitude, phase, and period values for the identified frequencies and their integer harmonics were subsequently calculated.

*3.1.1. Long-term Variations:* To investigate the long-term variability of DF Cyg, we analyzed the AAVSO, AFOEV, and *Kepler* datasets. The AAVSO data, spanning nearly 50 years (1968–2020), exhibit more than 20 long-term cycles. By calculating the mean brightness on a year-by-year and peak-to-peak basis, and avoiding short-term variation changes at $0^m.4$ and $2^m.08$, we determined the long-term period to be 0.00128197 $c/d$ (780.1 $\pm$ 1.1 d).

The *Kepler* light curves in Figure 1, reveal distinct long-term variations. All quarters are plotted. The brightness of DF Cyg varies between approximately 9.64 and 13.12 magnitudes over cycles that last around 773.9 $\pm$ 0.1 days. Each plot included three-quarters of the data in the subplots. In the first row of the multiplot (Q0-2 and Q9-11), the magnitude of DF Cyg is increased to a maximum brightness level ($\sim 9^m.64$), which takes approximately 140 and 290 days for Q0-2 and Q9-11, respectively. DF Cyg reaches its maximum brightness level and has a nearly constant brightness value ($\sim 10^m.3$) with respect to the long-term variations in the

---

[+] https://www.l3harrisgeospatial.com/Software-Technology/IDL
[*] A statistical technique for detecting periodic signals in unevenly sampled time series, particularly suited to identifying both short- and long-term variations.

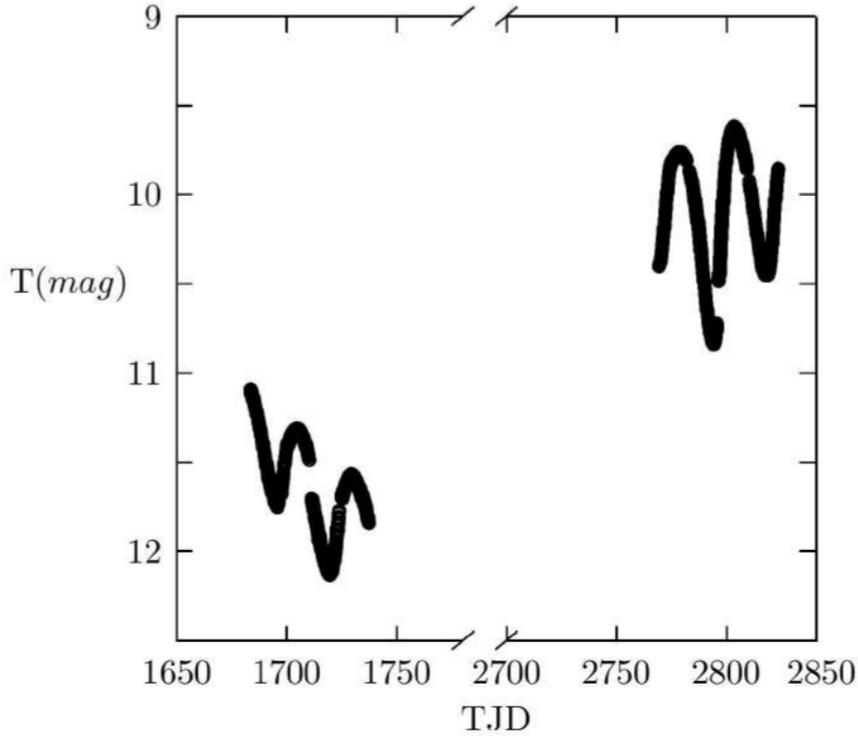

**Figure 3.** Derived light curve from sectors 14, 15, 54, and 55 observations of the *TESS* data. The time interval for the sectors 14, 15, 54, and 55 are approximately 1683-1710 TJD, 1710-1736 TJD, 2770-2796 TJD, and 2797-2824 TJD, respectively. TJD is equal to JD-2,457,000. The mean magnitude difference in the *TESS* filter band ($T$) associated with the long-term mean brightness variations is 0.009 $mag$ and it is 0.003 $mag$ between the sectors 14 and 15 and sectors 54 and 55.

second row of the multiplot (Q3-5 and Q12-14). The constant level continued for nearly 280 days and ∼275 days for Q3-5 and Q12-14, respectively. At Q6-8 and Q15-17, the brightness of DF Cyg decreased to the lowest level ($\sim 13^m.12$), as shown in the third row of Figure 1. This took ∼265 and ∼219 days for Q6-8 and Q15-17, respectively.

The feature of the light curve for long-term variations is mean brightness. The AFOEV data, spanning from 1927 to 2020, also show long-term variations with a period of 779.7 ± 1.8 days. The brightness of DF Cyg varied from $9^m.2$ to $14^m.5$ in AFOEV data. Despite the 40-year gap in the data, the mean brightness of the stars was 11.4 $mag$ and the overall trend was clear. The consistent long-term period measurements from the AAVSO, *Kepler*, and AFOEV data confirmed the robust nature of this variability and provided strong evidence for a long-term cycle in DF Cyg brightness. These findings are in agreement with those of previous studies (779.6 ± 0.2 days; Bódi et al. (2016); 775 days - Percy and Ursprung (2006); 795 ± 5 days - Vega, Stassun, Montez, Boyd and Somers (2017), and 784 ± 16 days - Manick et al. (2019)), with similar periods of approximately 775-795 days.

The *Kepler* and AAVSO data of DF Cyg revealed a difference in brightness of $3^m.50$ and $4^m.50$ between the long-term maximum and minimum values, respectively.

*3.1.2. Short-term Variations:* We performed a conventional Fourier analysis of all the data for DF Cyg. To precisely detect short-term variations, we removed long-term variations from the data as much as possible.

In our study, we addressed limitations in frequency analysis, including cadence and sampling strategy. We did this by carefully analyzing the spectral window function, checking for possible aliasing artifacts, and using *TESS* sectors with short cadences to overcome these challenges. We also carefully controlled the time interval and spectral frequency range of the sectors to assess the robustness of the identified frequencies and to distinguish true signals from sampling artifacts.

The *Kepler* data, spanning nearly four years, revealed approximately 29 alternating minimum cycles and 60 short-term cycles. As expected, the *Kepler* data were more accurate than the ground-based observations. The high-precision *Kepler* data allowed us to identify sub-harmonics and integer harmonics such as 2f, 3f, 4f, f/2, 3f/2, and 5f/4. For a detailed analysis, we continued to remove other variations to identify other sub-harmonics. The instrument sensitivity of the *Kepler* is almost one-third of the highest peak amplitude, which is our limit for the frequency analysis (Van Cleve and Caldwell, 2016). At this limit, the signal becomes lower than that of the three SNRs. All frequency outputs of the *Kepler* data depend on long- or short-term variations. In addition to frequency, we also analyzed amplitude variations of the fundamental period in cycle-to-cycle, and the *Kepler* data were cleaned to avoid long-term variations. The amplitude variations are related to the mean brightness change. The amplitude peak of each fundamental mode in each observation cycle varied between $10^m.763$ (Q11) and $11^m.631$ (Q17). Because of the high uncertainty that *Kepler* takes Q0 in the safe mode, we could not include the amplitude estimation even though the Q0 data had the lowest amplitude peak of the fundamental mode at $12^m.212$.

While AAVSO and AFOEV data offer long-term coverage, their lower precision and discontinuous nature, coupled with significant temporal variations, limit the detection of short-term variations. The resulting frequency spectra exhibited broadened peaks and splits because of the large timespans and temporal changes in the short-term AAVSO and AFOEV data. Due to these effects, only $f$ and $f/2$ become remarkable in the frequency spectra of AAVSO and only $f$ in AFOEV data, respectively. The other peaks are below the noise threshold.

The short-time-span data were gathered from different ground-based observations from ASAS, SuperWASP, and ASAS-SN to analyze short-term variations. The mean brightness of DF Cyg is detected as $11^m.806$ and $10^m.484$ in the V, and I bands in ASAS observations, respectively. The amplitude variations of the data were 3.41 $mag$ in the V-band and 2.87 $mag$ in the I-band. In the SuperWASP data, the mean brightness of DF Cyg was $12^m.465$. The mean brightness of the data in the V band is $12^m.03$, and the amplitude variation is $2^m.67$ in the ASAS-SN campaign data. All datasets have either discontinuous time intervals or low SNR to determine the subharmonics of the short-term frequencies. Thus, using these datasets, only $f$ and $f/2$ were significant.

The high-cadence *TESS* data, with a cadence of 120 s, offer unprecedented precision in capturing the short-term variability of DF Cyg. However, the limited time span of *TESS* observations restricts our ability to fully characterize the long-term behavior of stars.

**Table 2.** Derived short- and long-term pulsation frequencies (c/d) for DF Cyg, including their amplitudes (mag), phases (rad/$2\pi$), periods (days), and associated uncertainties. Ground- and space-based photometric datasets (*ASAS, ASAS-SN, AAVSO, SuperWASP, Kepler,* and *TESS*) are indicated under the "Observation" column. *TESS* sectors 14-15 and 54-55 labeled as TESS-S1415 and TESS-S5455, respectively. Uncertainties reflect $1\sigma$ confidence intervals.

| Frequency (c/d) | Amplitude (mag) | Phase (rad/$2\pi$) | Period (day) | Observation |
|---|---|---|---|---|
| 0.0405564 ± 2.84e-04 ($f$) | 0.3885 ± 2.84e-02 | 0.1923 ± 1.17e-02 | 24.658 ± 0.172 | ASAS-N |
| 0.0390912 ± 2.61e-04 ($f$) | 0.3630 ± 3.78e-02 | 0.4343 ± 1.81e-02 | 25.590 ± 0.174 | ASAS-SN |
| 0.0200098 ± 2.77e-04 ($f/2$) | 0.2029 ± 2.94e-02 | 0.3619 ± 2.31e-02 | 49.976 ± 0.691 | ASAS-SN |
| 0.0399514 ± 2.80e-04 ($f$) | 0.4422 ± 4.92e-02 | 0.4401 ± 1.83e-02 | 25.084 ± 0.180 | AAVSO |
| 0.0200096 ± 3.67e-04 ($f/2$) | 0.2876 ± 4.35e-02 | 0.2419 ± 2.40e-02 | 50.001 ± 0.931 | AAVSO |
| 0.0012820 ± 1.81e-06 ($f_{orb}$) | 1.0161 ± 6.16e-02 | 0.8639 ± 9.64e-03 | 780.041 ± 1.101 | AAVSO |
| 0.0012825 ± 2.94e-06 ($f_{orb}$) | 1.1370 ± 1.09e-02 | 0.9533 ± 1.53e-03 | 779.737 ± 1.785 | AFOEV |
| 0.0394424 ± 1.44e-04 ($f$) | 0.2220 ± 1.04e-02 | 0.1455 ± 7.48e-03 | 25.036 ± 0.090 | AFOEV |
| 0.0399665 ± 4.38e-04 ($f$) | 0.3720 ± 2.46e-02 | 0.7196 ± 1.05e-03 | 25.021 ± 0.274 | SuperWASP |
| 0.0210067 ± 8.69e-05 ($f/2$) | 0.1877 ± 2.46e-02 | 0.5303 ± 2.08e-03 | 47.604 ± 0.197 | SuperWASP |
| 0.0012921 ± 7.77e-08 ($f_{orb}$) | 1.2933 ± 2.69e-04 | 0.2689 ± 3.30e-05 | 773.928 ± 0.047 | Kepler |
| 0.0401234 ± 2.53e-07 ($f$) | 0.4118 ± 2.69e-04 | 0.4469 ± 1.08e-04 | 24.923 ± 0.002 | Kepler |
| 0.0200277 ± 1.19e-06 ($f/2$) | 0.0893 ± 2.69e-04 | 0.3852 ± 5.07e-04 | 49.931 ± 0.003 | Kepler |
| 0.0802469 ± 1.09e-06 ($2f$) | 0.9166 ± 2.69e-04 | 0.4961 ± 4.63e-04 | 12.462 ± 0.002 | Kepler |
| 0.0601512 ± 1.97e-06 ($3f/2$) | 0.4828 ± 2.69e-04 | 0.3883 ± 8.38e-04 | 16.625 ± 0.001 | Kepler |
| 0.0135332 ± 4.86e-06 ($f/3$) | 0.1898 ± 2.69e-04 | 0.3972 ± 2.06e-03 | 73.893 ± 0.027 | Kepler |
| 0.0105749 ± 3.61e-06 ($f/4$) | 0.2242 ± 2.69e-04 | 0.1486 ± 1.53e-03 | 94.563 ± 0.032 | Kepler |
| 0.0486922 ± 6.46e-06 ($5f/4$) | 0.1680 ± 2.69e-04 | 0.6064 ± 2.74e-03 | 20.537 ± 0.003 | Kepler |
| 0.0888547 ± 5.05e-05 ($2f$) | 0.0159 ± 7.84e-05 | 0.3993 ± 7.87e-04 | 11.254 ± 0.006 | TESS-S1415 |
| 0.1351333 ± 1.49e-04 ($3f$) | 0.0054 ± 7.84e-05 | 0.4766 ± 2.32e-03 | 7.400 ± 0.008 | TESS-S1415 |
| 0.1684538 ± 8.72e-05 ($4f$) | 0.0092 ± 7.84e-05 | 0.4603 ± 1.36e-03 | 5.936 ± 0.003 | TESS-S1415 |
| 0.2202857 ± 2.09e-04 ($5f$) | 0.0038 ± 7.84e-05 | 0.1842 ± 3.25e-03 | 4.540 ± 0.004 | TESS-S1415 |
| 0.2730432 ± 2.19e-04 ($6f$) | 0.0037 ± 7.84e-05 | 0.6679 ± 3.41e-03 | 3.662 ± 0.003 | TESS-S1415 |
| 0.3406098 ± 5.95e-04 ($7f$) | 0.0013 ± 7.84e-05 | 0.9400 ± 9.28e-03 | 2.936 ± 0.005 | TESS-S1415 |
| 0.3693025 ± 3.33e-04 ($8f$) | 0.0024 ± 7.84e-05 | 0.8290 ± 5.20e-03 | 2.708 ± 0.002 | TESS-S1415 |
| 0.0791108 ± 2.71e-06 ($2f$) | 0.1053 ± 2.93e-05 | 0.8276 ± 4.26e-05 | 12.641 ± 0.001 | TESS-S5455 |
| 0.1766196 ± 4.39e-05 ($2f/9$) | 0.0031 ± 2.93e-05 | 0.2168 ± 6.89e-04 | 5.662 ± 0.003 | TESS-S5455 |
| 0.1333846 ± 3.45e-05 ($3f$) | 0.0065 ± 2.93e-05 | 0.5656 ± 5.41e-04 | 7.497 ± 0.004 | TESS-S5455 |
| 0.1609814 ± 1.99e-05 ($4f$) | 0.0178 ± 2.93e-05 | 0.9584 ± 3.12e-04 | 6.212 ± 0.002 | TESS-S5455 |
| 0.2042164 ± 3.05e-05 ($5f$) | 0.0094 ± 2.93e-05 | 0.6484 ± 4.78e-04 | 4.897 ± 0.002 | TESS-S5455 |
| 0.2732084 ± 7.88e-05 ($6f$) | 0.0033 ± 2.93e-05 | 0.4205 ± 1.24e-03 | 3.660 ± 0.002 | TESS-S5455 |
| 0.2364126 ± 5.76e-05 ($7f$) | 0.0048 ± 2.93e-05 | 0.4039 ± 9.04e-04 | 4.230 ± 0.002 | TESS-S5455 |
| 0.3146036 ± 6.83e-05 ($8f$) | 0.0027 ± 2.93e-05 | 0.6792 ± 1.07e-03 | 3.179 ± 0.001 | TESS-S5455 |

Despite this limitation, the *TESS* data provided valuable insights into the rapid variations and pulsational modes of the DF Cyg. In this study, we employed *TESS* data for the first time. In contrast to *Kepler* data, *TESS* observations of DF Cyg have 120-second cadence (in sectors 15, 54, and 55) data and are allowed to investigate short-term variations in much more detail. With a high cadence rate, the most precise data in this study was *TESS*. However, as stated above, the data covers only ~100 d. The light curve of the *TESS* observations has only three and a half minimum cycles with ~9800 data points (Figure 3). The light curve of the target, derived from sectors 14 and 15, had two alternating short-term cycles. It should be noted that the accuracy of the Sector 14 data is nearly equal to that of the *Kepler* data, but sectors 15, 54, and 55 that were observed in the 120 s cadence mode are the most precise observation data to date for the DF Cyg. From Sector 14 to Sector 15 observations of *TESS*, the brightness of DF Cyg decreased, similar to the Q6-8 and Q15-17 data of the *Kepler* observations ( third row of Figure 1). However, the brightness of the star remained almost constant in sectors 54 and 55, as in the *Kepler* Q3-5 and Q12-Q14 data ( second row of Figure 1). The mean magnitude difference in the *TESS* filter band ($T$) regarding long-term mean brightness variations is 0.009 $mag$ and 0.003 $mag$ between sectors 14 and 15, and sectors 54 and 55 of the *TESS* observations.

The results obtained from the analysis of *TESS* observations of DF Cyg in this study identified low-amplitude harmonics and subharmonics that were not accessible in previous studies. We achieved this by avoiding situations where aliasing could mask or mimic real signals when interpreting frequencies close to the Nyquist limit, thanks to the sensitivity of the short-cadence *TESS* data. Using *TESS* data and removing long-term variations from the data, we estimated short-term variations in the fundamental frequency ($f$) of pulsation and its integer harmonics. Because of the two alternating short-term cycles of the *TESS* light curve, we clearly detected $f, f/3, 2f, 2f/9, 3f, 3f/2, 4f, 5f, 6f, 7f$ and $8f$ in the *TESS* power spectrum of the DF Cyg.

To further investigate the frequency of DF Cyg variability, we performed a comprehensive periodogram analysis of the combined space- and ground-based photometric data. The analysis revealed 33 fundamental frequencies and orbital periods from space- and ground-based photometric data. Table 2 lists these frequencies, along with their corresponding amplitudes, phases, and estimated periods with uncertainties.

The long- and short-term periods derived from the various datasets (AAVSO, AFOEV, *Kepler*, and *TESS*) are remarkably consistent. Minor discrepancies in the frequency values can be attributed to inherent differences in the data quality, time span, and sampling intervals.

*3.2. Spectroscopic Analysis*

To gain further insight into the physical processes driving DF Cyg's variability, we supplemented our photometric analysis with spectroscopic observations. By examining the detailed line profiles and radial velocity variations, we aimed to better understand the atmospheric properties of star.

Our spectroscopic data served two main purposes. First, we employed the spectra to measure radial velocities (RVs) through cross-correlation analysis. This method enabled us to determine the phase-dependent kinematics of the star, which are directly linked to its pulsation (see Table 3 and Figure 4). Second, the spectra were essential for deriving the star's fundamental atmospheric parameters: effective temperature ($T_{\text{eff}}$), surface gravity ($\log g$), and metallicity ($[Fe/H]$). We obtained these parameters by conducting a detailed spectral fitting with the Stellar Parameters and Chemical Abundance Estimator (SP_Ace) code (Boeche and Grebel, 2016) and refining the results using a Markov Chain Monte Carlo (MCMC) approach, as detailed in Tables 4 and 5. The resulting atmospheric parameters were then used to constrain the star's physical properties for our SED modeling and to place the star in an evolutionary context. It is important to note that, compared to previous spectroscopic analyses of DF Cyg, our study benefits from a careful consideration of the effects of stellar variability on the spectral line profiles. We acknowledge that a detailed chemical abundance analysis was not feasible due to line-profile distortions caused by shock phenomena and the limited signal-to-noise ratio (SNR) of the spectra (see Section 3.2). Therefore, the spectra were used as diagnostic tools for atmospheric parameters and RVs, but not for chemical abundances.

**Table 3.** RV data derived from the McDonald spectra of the DF Cyg. The phase of short-term variations was calculated using the minimum time of short-term variations as $T_0 = 2454293.331$ (HJD) and the period of short-term variations as $P = 24.9$ days.

| Time (HJD-2450000) | Phase | RV (km s$^{-1}$) |
|---|---|---|
| 4726.66991635 | 0.33 | -40.33 |
| 4728.65926948 | 0.41 | -37.96 |
| 4759.64179477 | 0.65 | -3.45 |
| 4787.65328437 | 0.77 | 1.09 |

To mitigate the impact of line profile variations, we employed a cross-correlation function (CCF♯) analysis to determine the radial velocity. CCF provides a more robust measure of the Doppler shift, as it considers the overall line profile rather than the individual line features. We obtained individual RV data from the observed spectra of the DF Cyg using the cross-correlation method (Allende Prieto, 2007). RV data were derived using a selected Atlas of Arcturus atomic line mask with the ISPEC code (Blanco-Cuaresma, 2019). From the cross-correlation profile, the RV data were determined from the Gaussian fitted mean line despite the low SNR. The derived radial velocities are presented in Table 3.

Figures C1, C2, and C3 show examples of DF Cyg spectra recorded at HJD 2454726.66991635 (September 16, 2008), HJD 2454759.64179477 (October 19, 2008), and HJD 2454787.65328437 (November 16, 2008) in different wavelength ranges, along with the smoothed observed spectrum of the star to help guide the line profiles, highlighting the presence of line-profile distortions, particularly in the spectrum observed on November 16, 2008. These distortions are likely to be associated with shock-induced phenomena (e.g., line-doubling and line distortions). A detailed analysis of the line profiles, including the

---

♯ The CCF, with the textbook definition, is a Voigt profile centered on a radial velocity that records only the orbital motion and pulsation during the phases when the stellar atmosphere was stable.

identification of affected lines and quantification of their distortions, would provide valuable insights into the physical processes driving the variability of DF Cyg. However, such an analysis requires high-quality spectra with a sufficient signal-to-noise ratio and spectral resolution. In Table D1, we provide a list of Fe I and Fe II that exhibit such distortions in the McDonald spectra of DF Cyg. These iron lines were partially tested by (Şahin et al., 2011; Şahin and Bilir, 2020; Şahin et al., 2023; Marışmak et al., 2024; Şahin et al., 2024; Sentürk et al., 2024; Cınar et al., 2025) to derive solar abundances have been used to assess uncertainties in atomic data (e.g., $\log gf$). The lines appear to be the weakest in the spectrum of the star recorded at HJD 2454787.65328437 (November 16, 2008). However, these lines show smooth profiles (the lines are easily recognizable because of the relatively high SNR) in the spectrum observed on September 16. This is because the spectrum of DF Cyg observed on September 16 was obtained in a brighter phase ($\sim$ 12 mag.) than other spectra taken in the long term, and at the 0.3 phase in the short-term variation. In the McDonald spectra obtained on October 19 and November 16 there was no measurable isolated line to determine the model atmospheric parameters of DF Cyg using the classical spectroscopic analysis method (i.e., ionization and excitation equilibria of Fe lines).

The cross-correlation function (CCF) profiles of the spectra were examined to determine which of the four spectra might be preferred for abundance analysis. Notably, the CCF profiles exhibited phase-dependent morphologies linked to pulsation cycles (Figure 4). During pulsation-driven atmospheric expansion, double-peaked profiles emerged, while quiescent phases ($\phi \sim 0.77$) produced symmetric, single-peaked profiles. A similar pulsation effect on the estimated CCF profile was reported by Manick et al. (2019). The November 16, 2008 spectrum†† (HJD 2454787.65328437), acquired during a stable phase, was prioritized for atmospheric modeling.

To the best of our knowledge, the available studies for the star based on (classical) spectroscopic analysis are limited by Giridhar et al. (2005), and Mohorian et al. (2024). Giridhar et al. (2005) analyzed the echelle spectrum of the star and reported the abundances of several key elements (13 elements in total), most of which were represented by only a few lines (e.g., one line for Ti I and C I, and two lines for Cr II). The model parameters reported by Giridhar et al. (2005) were $T_{\text{eff}} = 4800$ K, $\log g = 1.7$ cm s$^{-2}$, $[Fe/H] = 0.0$ dex, and $v_{\text{mic}} = 3.6$ km s$^{-1}$. Surprisingly, the echelle spectra (taken at the $\sim$0.84 phase) by Giridhar et al. (2005) were found to have a nearly identical phase ($\phi = 0.77$) to the McDonald spectra obtained on November 16, 2008. Owing to the lack of spectral lines with clear measurable profiles for EW analysis, we preferred spectrum fitting to determine the atmospheric parameters of DF Cyg. This technique allows us to obtain the best match between the observed spectra and spectral models. For this purpose, we used SP_Ace code. SP_Ace constructs a spectral model in real time while performing spectral analysis. This is achieved by retrieving information about the central wavelengths and strengths of the spectral lines from the embedded Generalized Curve of Growth (GCOG) library prepared specifically for the code, which is used to assume a line profile and generate a model spectrum. The spectrum-fitting technique provided model parameters in excellent agreement with those reported by Giridhar et al. (2005), as follows: $T_{\text{eff}} = 4781^{+66}_{-63}$ K, $\log g = 1.74^{+0.20}_{-0.19}$ cm s$^{-2}$, and [Fe/H] = $-0.07^{+0.07}_{-0.09}$ dex. Because SP_Ace does not provide a micro-turbulent velocity, it is calculated using Equation 1, where $T_0 = 5500$ K and $g_0 = 4.0$ dex, provided by (Blanco-Cuaresma, 2019) to analyze FGK stars in the ISPEC code as $v_{\text{mic}} = 1.4$ km s$^{-1}$. Figure 5 shows the observed and fitted model spectra produced using the SP_Ace.

---

†† It is exhibited a single-peaked (indicating shock-free stellar atmospheric dynamics) CCF profile.

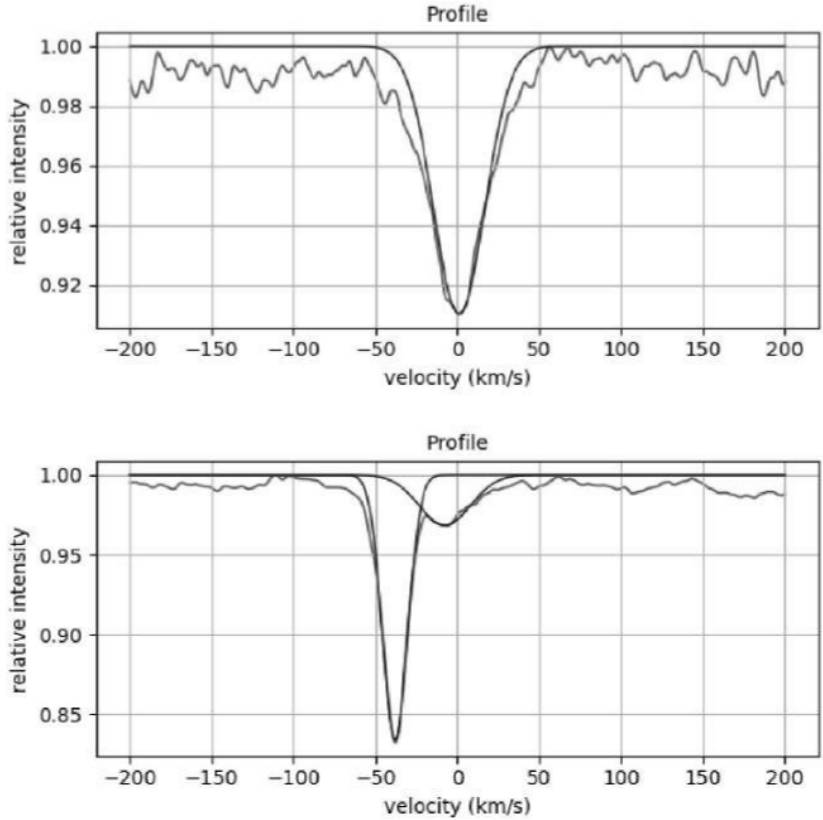

**Figure 4.** Estimated CCF profile from DF Cyg McDonald spectra (2454787.65328437 HJD at $\phi \sim 0.77$ and 2454726.66991635 HJD at $\phi \sim 0.33$, respectively). The CCF measurements were computed using ISPEC code. The shape of the CCF profile depended on the pulsation phase. When the star is at the "normal" radius, the CCF profile velocity is nearly zero, and the profile returns Symmetric shapes (above). However, when the star is pulsating, the CCF line profile becomes a two-peak shape because the atmosphere of the star is unstable (below).

While the signal-to-noise ratio (SNR) of our spectra peaks at $\sim 20$ (Section 2.2), which is suboptimal for classical equivalent-width abundance analysis, the SP_Ace spectral-fitting technique remains robust for estimating global atmospheric parameters ($T_{\text{eff}}, \log g, [Fe/H]$) at this SNR level. The uncertainties reported in Table 5 ($\Delta \log g \approx \pm$ 0.25–0.55 dex; $\Delta [Fe/H] \approx \pm$ 0.18–0.31 dex) reflect these limitations. Consistency with independent results from Giridhar et al. (2005) ($T_{\text{eff}}$=4800) K, $\log g$=1.7 dex, $[Fe/H]$=0.0 dex) supports the reliability of our estimates for diagnostic purposes, though higher-SNR spectra would be required for precise chemical abundance analysis.

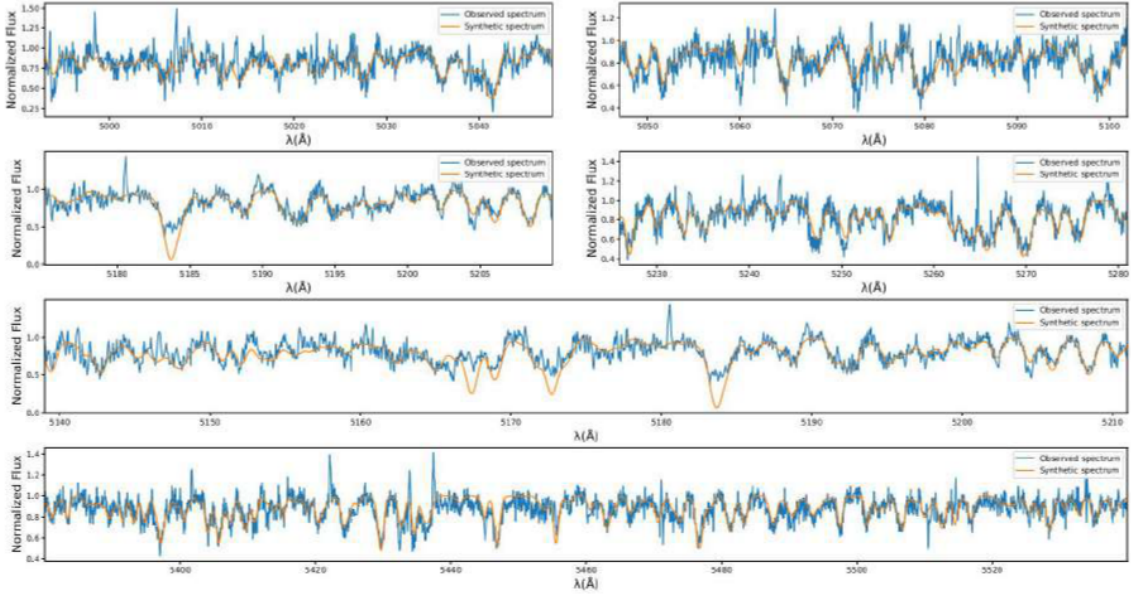

**Figure 5.** The McDonald spectra of DF Cyg obtained on HJD 2454787.65328437 (November 16, 2008; $\phi \sim 0.7$) for several separate wavelength regions. The SP_Ace computed synthetic spectrum for $T_{\text{eff}} = 4781$ K, $\log g = 1.74$ cm s$^{-2}$, and [Fe/H] = -0.07 dex is also presented.

$$\begin{aligned} v_{\text{mic}} =& 1.25 + 4.01\text{e-}4(T_{\text{eff}} - T_0) + 3.1\text{e-}7(T_{\text{eff}} - T_0)^2 \\ & - 0.14(log g - g_0) - 0.005(log g - g_0)^2 \\ & + 0.05[Fe/H] + 0.01[Fe/H]^2 \end{aligned} \quad (1)$$

To further validate our results, we reanalyzed the *HERMES* spectra (Maksym Mohorian, 2024, private communication) of DF Cyg with visit IDs 83 (972481) and 26 (412205) listed in Mohorian et al. (2024) using SP_Ace code. Our analysis of the *HERMES* spectrum with visit ID 83 yielded $T_{\text{eff}} = 5657$ K, $\log g = 2.04$ cm s$^{-2}$, $[Fe/H] = 0.25$ dex, and $v_{\text{mic}} = 3.6$ km s$^{-1}$. For visit ID 26, the parameters were $T_{\text{eff}} = 5707$ K, $\log g = 1.56$ cm s$^{-2}$, $[Fe/H] = 0.28$ dex, and $v_{\text{mic}} = 3.6$ km s$^{-1}$. These values differ slightly from those reported by Mohorian et al. (2024): $\Delta T_{\text{eff}} = -113$ K, $\Delta \log g = 0.12$ dex, $\Delta[Fe/H]) = 0.2$ dex for visit ID 83 and $\Delta T_{\text{eff}} = -43$ K, $\Delta \log g = -0.15$ dex, and $\Delta[Fe/H]) = 0.23$ dex for visit ID 26. These small discrepancies validate the applicability of the SP_Ace code for analyzing these spectra, and highlight the importance of careful spectral analysis and the potential impact of different analysis techniques.

To further refine our analysis and assess the uncertainties in the derived parameters, we utilized Markov Chain Monte Carlo (MCMC) techniques. We employed the ISPECMCMC framework in Python, which incorporates the ispec backend (Blanco-Cuaresma et al., 2014) and emcee ensemble samplers (Foreman-Mackey et al., 2013). This framework was developed for the optimization of the seven spectral parameters (effective temperature – $T_{\text{eff}}$, log surface gravity – $\log g$, metallicity –[M/H], rotational velocity –vsini, microturbulence velocity – $v_{\text{mic}}$, Doppler shift correction velocity – dop(a radial velocity correction to spectral lines in the spectrum due to Doppler shift), and vertical shift velocity $v_{\text{shift}}$ (a free parameter used to smooth out low velocity shifts in flux due to the normalization of each spectrum) of DF Cyg. By exploring the parameter space and accounting for uncertainties in the observational data, we obtained robust estimates of the atmospheric parameters. The optimization process was limited to the regions that were determined using the selected Fe lines(The atomic reliability of these lines was tested using the solar and F stars spectra in Şahin and Bilir (2020)) from Şahin and Bilir (2020). Spectral parameter priors were sampled with a normal distribution of approximately 10% for a good fit. Using `emcee`, which is a built-in autocorrelation time module, we achieved a sufficient number of iterations and ensured convergence. This method was applied to all spectral data with a total of 200,000 iterations using 64 walkers to obtain the final solutions for each spectrum (Figs. A1, A2, A3, and A4). We present our optimal parameters with uncertainties for each spectral parameter estimated from the 16th, 50th, and 84th percentiles of each distribution from the Markov chain MCMC results for the posterior probability distributions listed in Table 4. We also show correlations between the parameters on a corner plot for each spectral data point in the figures A1, A2, A3, and A4. The corner plots show $\chi^2$ distributions of $T_{\text{eff}}$, $\log g$, vsini, $[M/H]$, $v_{\text{mic}}$, dop, and $v_{\text{shift}}$. The solid contours represent the densities of posterior distributions. The histogram distributions, represented by solid lines, are plotted across quantities at the 16th, 50th, and 84th percentiles with dashed lines. The model parameters from the MCMC analysis provided in Table 5, which are based on the Fe lines, are compatible with those from the spectrum fitting results by SP_Ace and those of Giridhar et al. (2005), within the uncertainties of the model parameters.

**Table 4.** The best optimal parameters with uncertainties of DF Cyg on effective temperature ($T_{\text{eff}}$), logarithmic surface gravity ($\log g$), metallicity ($[M/H]$), rotational velocity ($v\sin i$), microturbulence velocity ($v_{\text{mic}}$), Doppler shift correction velocity (dop), and vertical shift velocity ($v_{\text{shift}}$) based on the MCMC results for each observed spectrum are listed. $T_{\text{eff}}$ is given in K unit, $\log g$ and $[M/H]$ are in dex unit, and $v\sin i$, $v_{\text{mic}}$, dop and $v_{\text{shift}}$ are in km s$^{-1}$ unit.

| Time (HJD-2450000) | $T_{\text{eff}}$ (K) | $\log g$ (cm s$^{-2}$) | $[M/H]$ (dex) | $v\sin i$ (km s$^{-1}$) | $v_{\text{mic}}$ (km s$^{-1}$) | dop (km s$^{-1}$) | $v_{\text{shift}}$ (km s$^{-1}$) |
|---|---|---|---|---|---|---|---|
| 4726.66991635 | $5441.67^{+393.40}_{-399.37}$ | $2.15^{+1.04}_{-1.02}$ | $-0.37^{+0.34}_{-0.39}$ | $18.95^{+5.01}_{-4.06}$ | $1.92^{+1.43}_{-1.34}$ | $0.84^{+1.57}_{-2.42}$ | $-0.05^{+0.02}_{-0.02}$ |
| 4728.65926948 | $5316.06^{+344.81}_{-342.84}$ | $2.22^{+0.93}_{-0.91}$ | $-0.15^{+0.29}_{-0.32}$ | $16.20^{+3.50}_{-3.00}$ | $1.99^{+1.37}_{-1.35}$ | $0.47^{+1.34}_{-1.33}$ | $-0.03^{+0.02}_{-0.02}$ |
| 4759.64179477 | $5106.22^{+409.68}_{-402.72}$ | $1.78^{+1.00}_{-0.94}$ | $-0.46^{+0.39}_{-0.44}$ | $26.25^{+6.21}_{-4.80}$ | $2.04^{+1.32}_{-1.38}$ | $-2.31^{+2.37}_{-1.67}$ | $-0.04^{+0.03}_{-0.03}$ |
| 4787.65328437 | $5015.42^{+526.56}_{-470.54}$ | $1.89^{+1.11}_{-1.03}$ | $-0.49^{+0.46}_{-0.53}$ | $35.74^{+7.98}_{-6.45}$ | $2.05^{+1.33}_{-1.41}$ | $1.99^{+3.36}_{-3.20}$ | $-0.05^{+0.03}_{-0.02}$ |

### 3.3. SED Analysis

We used Virtual Observatory SED Analyzer (VOSA) tool (Bayo et al., 2008)) and spectrAl eneRgy dIstribution bAyesian moDel averagiNg fittEr (`ARIADNE`) (Vines and Jenkins, 2022) code to analyze the spectral energy distribution (SED) of DF Cyg. We selected photometric observations of the star in the broad wavelength range between ~3500 and ~52 000 Å that were used with synthetic stellar atmospheric models (Kurucz, 1993; Castelli and Kurucz, 2003; Husser et al., 2013; Allard et al., 2012) to fit the shape of the stellar SED. Fortunately, DF Cyg was observed in broad-band photometry bandpasses (*2MASS JHKs; Johnson UBV; Cousins IR; WISE W1, W2, W3 and W4; PS1 griyz; SDSS ugriz; TYCHO BV; Gaia DR3 G,BP and RP; Kepler; TESS; AKARI IRC.S9W, IRC.L18W and IRAS 12,25, 60 and 100 µs*). Photometric data of DF Cyg were obtained from the VizieR, MAST, and Gaia DR3 databases using `astroquery`.

**Table 5.** Same as Table 4 but obtained using neutral and ionised iron lines (Fe I and Fe II) of DF Cyg spectra.

| Time (HJD-2450000) | $T_{\text{eff}}$ (K) | $\log g$ (cm s$^{-2}$) | [M/H] (dex) | $v \sin i$ (km s$^{-1}$) | $v_{\text{mic}}$ (km s$^{-1}$) | dop (km s$^{-1}$) | $v_{\text{shift}}$ (km s$^{-1}$) |
|---|---|---|---|---|---|---|---|
| 4726.66991635 | $5045.46^{+228.48}_{-227.32}$ | $1.15^{+0.25}_{-0.43}$ | $-0.24^{+0.18}_{-0.20}$ | $16.82^{+2.59}_{-2.36}$ | $1.91^{+1.46}_{-1.33}$ | $0.39^{+0.71}_{-0.64}$ | $-0.01^{+0.01}_{-0.01}$ |
| 4728.65926948 | $5216.38^{+242.71}_{-256.52}$ | $1.07^{+0.31}_{-0.46}$ | $-0.24^{+0.18}_{-0.21}$ | $21.33^{+4.21}_{-3.51}$ | $1.88^{+1.42}_{-1.29}$ | $0.57^{+1.27}_{-2.28}$ | $-0.01^{+0.01}_{-0.01}$ |
| 4759.64179477 | $4932.19^{+263.06}_{-280.63}$ | $1.02^{+0.34}_{-0.55}$ | $-0.35^{+0.21}_{-0.31}$ | $28.42^{+4.27}_{-4.02}$ | $2.09^{+1.24}_{-1.34}$ | $-2.55^{+1.27}_{-1.25}$ | $-0.01^{+0.01}_{-0.01}$ |
| 4787.65328437 | $4732.67^{+301.76}_{-280.61}$ | $0.91^{+0.41}_{-0.51}$ | $-0.39^{+0.25}_{-0.31}$ | $37.79^{+5.64}_{-4.69}$ | $1.84^{+1.47}_{-1.27}$ | $0.89^{+2.45}_{-2.08}$ | $-0.01^{+0.01}_{-0.01}$ |

To model the observed SED, we employed various stellar atmosphere models: *Blackbody* and Castelli & Kurucz (Castelli and Kurucz, 2003, hereafter C&K2003), Kurucz (Kurucz, 1993), PHOENIX v2 (Husser et al., 2013) and BT-Settl (Allard et al., 2012).

Both VOSA and ARIADNE were used to perform Bayesian analysis and estimate the following stellar parameters: bolometric luminosity ($L_{\text{bol}}$), $T_{\text{eff}}$, $\log g$, and [M/H]. VOSA utilizes a $\chi^2$ minimization approach, whereas `ARIADNE` employs a nested sampling algorithm for Bayesian inference. Initial estimates for $T_{\text{eff}}$, $\log g$, and [M/H] were obtained from the spectroscopic analysis of DF Cyg in this study and were used as starting points for SED fitting.

Interstellar extinction was accounted for using $A_V = 0.6$ mag (Vega, Montez, Stassun and Boyd, 2017) and E(B-V) = 0.22 mag (Dawson, 1979) for VOSA (Verhoelst et al. 2017) and the extinction package by Barbary (2016) with the Fitzpatrick (1999) in `ARIADNE`.

To accurately model the spectral energy distribution (SED) of DF Cyg, we divided the photometric data into two distinct groups based on the long-term brightness variations observed in the light curve (Figure 1). These two groups, denoted SED-ID1 and SED-ID2, represent distinct phases of the star's long-term variability.

(i) SED-ID1 includes Johnson, Tycho, and SDSS photometry bandpasses, increases to the maximum brightness level.

(ii) SED-ID2 includes *Gaia* DR3, PS1, and *TESS* photometry bandpasses. It has a brightness of DF Cyg, which decreases to the lowest level of long-term variation.

Unfortunately, owing to the limited observational coverage at extreme brightness phases, some data, such as 2MASS (Skrutskie et al., 2006), were excluded from the SED analyses.

To complement the optical and near-infrared data, we incorporated observations from WISE, AKARI, and IRAS to capture infrared and radio emissions from DF Cyg. A prominent infrared excess is evident in both SED-ID1 and SED-ID2, indicating the presence of circumstellar dust (Figure 6). By analyzing the double-peaked structure of the SED, we determined that the peak wavelength of the dust emission was $\log(\lambda) = 4.523$ Å. Using Wien's displacement law, we estimated the temperature of dust to be approximately 869 K.

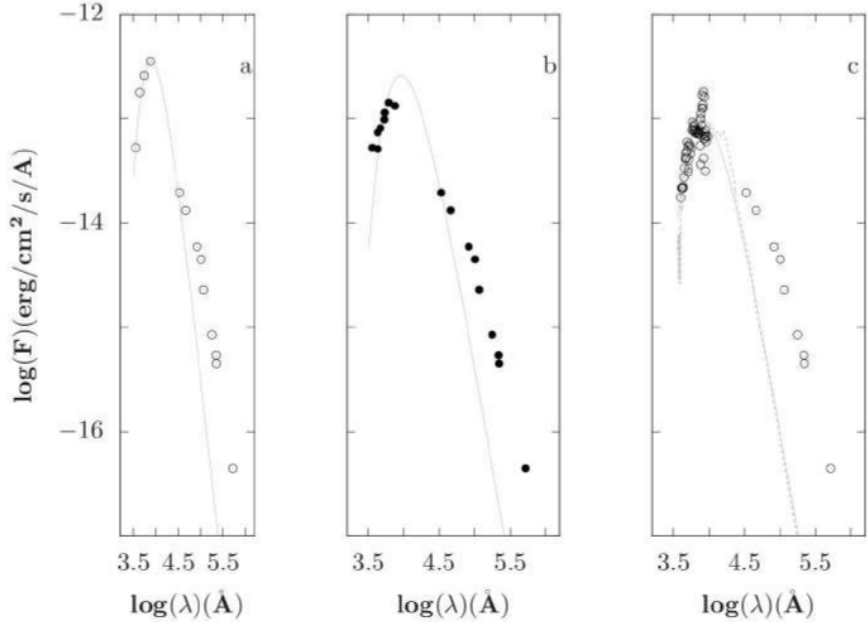

**Figure 6.** VOSA SED models and observations of DF Cyg. SED-ID1 and 2 with infrared observation data are shown in *panels a and b*, respectively. *Panel c* is included all observed data in different bandpasses in every phase of the long-term period of the star. The gray and dotted lines in the panels represent SED models obtained using the *Blackbody* and C&K2003 synthetic stellar atmospheric models, respectively. Broad infrared excess of the star is clearly seen in all panels.

**Table 6.** The `ARIADNE` SED fitting result parameters ($T_{\rm eff}$, $\log g$ and $[Fe/H]$) for SED-ID1 and 2 for DF Cyg, and their uncertainties are listed. $T_{\rm eff}$ is given in K units, $\log g$ and $[Fe/H]$ are in dex units.

|  | SED-ID1 | SED-ID2 |
|---|---|---|
| $T_{\rm eff}$ (K) | $5156.63^{+140.06}_{-148.90}$ | $5360.60^{+56.17}_{-47.05}$ |
| $\log g$ (cm s$^{-2}$) | $1.91^{+0.15}_{-0.15}$ | $2.00^{+0.05}_{-0.04}$ |
| $[Fe/H]$ (dex) | $-0.32^{+0.10}_{-0.10}$ | $-0.32^{+0.06}_{-0.05}$ |

The SEDs for each subgroup are shown in Figure 6 along with the observed fluxes. Scattering is evident in the observed fluxes in panel *c* (Figure 6) is likely due to the stellar pulsation.

SED-IDs 1 and 2 were modeled using a Blackbody function, whereas SED-ID 3 was fitted with both Blackbody and C&K2003 models in VOSA. However, the limited number of observational data points for DF Cyg precluded reliable determination of stellar atmospheric parameters using VOSA. Additionally, VOSA's lack of prior information and interpolation of model grid points, which are crucial for SED shape-independent parameters (e.g., $[M/H]$ and log $g$), necessitated the use of `ARIADNE` with Bayesian Model Averaging for SED-IDs 1 and 2.

Dynamic nested sampling with 100,000 iterations and 64 threads was employed to estimate the fitting parameters and their uncertainties. The parameters derived for DF and Cyg from the SED fitting models (Figs. B1 and B2) are listed in Table 6. The corresponding parameter distributions are shown in Figure B3 and B4.

DF Cyg (Gaia DR3 2078705244312711040) was observed during the *Gaia* space mission, and its parallax was obtained from the third data release (Gaia DR3) as 0.2746 ± 0.0236 *mas* (Gaia Collaboration et al., 2023). Leveraging C&K2003 synthetic stellar atmospheric models, the *Gaia* parallax, and the best-fit interstellar extinction ($A_v$) from `ARIADNE`, the interstellar extinctions for SED-IDs 1 and 2 were estimated to be $0.61^{+0.05}_{-0.05}$ and $0.60^{+0.01}_{-0.00}$, respectively.

The stellar radii for SED-IDs 1 and 2 were determined to be $33.615^{+3.899}_{-2.808}$ $R_\odot$ and $16.610^{+0.621}_{-0.613}$ $R_\odot$, respectively ( Figs. B3 and B4).

A comparison of the MCMC results (Table 4) with the SED results for DF Cyg (Table 6) revealed satisfactory agreement in terms of $T_{\text{eff}}$, $\log g$, and $[Fe/H]$. A detailed discussion is provided in Section 5.

Gezer et al. (2015) established a correlation between disk sources and binarity in confirmed binary RV Tauri stars. Similarly, Vega, Stassun, Montez, Boyd and Somers (2017) analyzed *Kepler* observations of DF Cyg, and linked long-term variations to binarity and disk obscuration. In our study, we also identified a pronounced infrared excess in the SED, which is indicative of a disk (Figure 6). Similar to other RV Tauri stars with disks, (Van Winckel, 2003; Kamath et al., 2014), DF Cyg exhibited double-peaked SED. Using Wien's displacement law, we estimated a disk temperature of ∼870 K, which was 100 K higher than the value reported by Vega, Stassun, Montez, Boyd and Somers (2017). Corporaal et al. (2023) provided evidence of circumbinary disks around evolved stars by using mid-infrared interferometric data. While we detected a disk around the DF Cyg, further analysis was limited, preventing us from deriving detailed structural information.

## 4. Period-Luminosity and Period-Radius Relations

RV Tauri stars, being Type II Cepheids, are amenable to period-luminosity (PL) and period-radius (PR) relations calibrated by large sky surveys (Matsunaga et al., 2009; Ripepi et al., 2015; Groenewegen and Jurkovic, 2017; Ripepi et al., 2019; Ngeow et al., 2022; Jurkovic et al., 2023). In this study, we employ the PL and PR relations presented in Ngeow et al. (2022) and Jurkovic et al. (2023), respectively, to derive the luminosity and radius of DF Cyg. We adopted the classical form of the PR, we assumed that the dominant pulsation mode corresponded to the fundamental radial mode and that the star was spherically symmetric. We neglected stellar rotation, magnetic field effects, and non-radial modes in our calculations. The resulting stellar luminosity and radius were determined as $619 \pm 195$ $L_\odot$ and $36.53 \pm 1.39$ $R_\odot$, respectively. Notably, the stellar radius and luminosity derived from the PR and PL relations, respectively, exhibited remarkable agreement with those obtained from SED

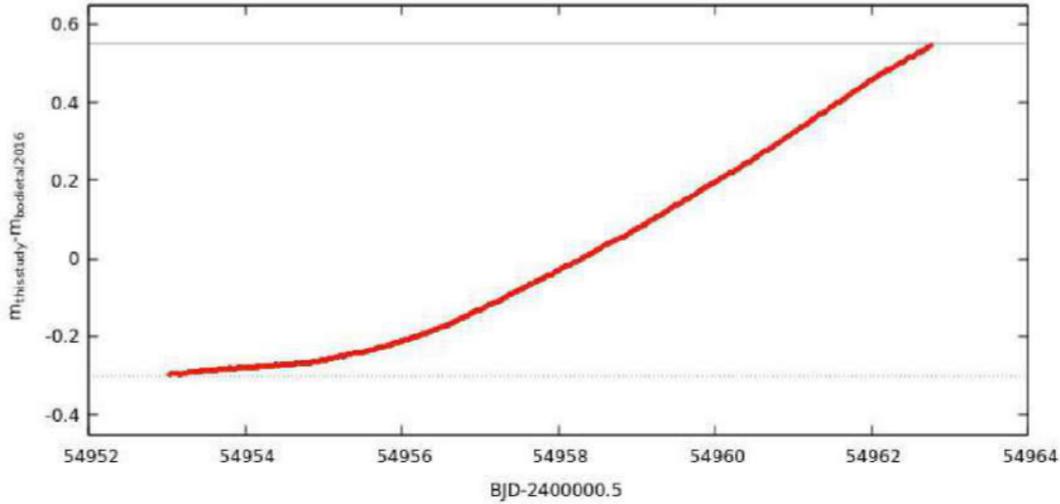

**Figure 7.** Brightness difference between derived light curves from Q0 *Kepler* data employed in this study and in Bódi et al. (2016) for DF Cyg. The difference comes from the fact that Bódi et al. (2016) used the raw light curve for DF Cyg. The dotted and solid lines represent $\Delta m = -0.30$ mag and $\Delta m = 0.55$ mag, respectively.

analysis incorporating Gaia DR3 parallax (see Figure B3).

In their study, Manick et al. (2019) employed a PLC relation developed using data from the OGLE-III survey of RV Tauri stars in the Large Magellan Cloud to estimate the luminosity of DF Cyg as $990 \pm 190$ L$_\odot$. However, in this study, the luminosity of DF Cyg was obtained as $619 \pm 195$ L$_\odot$ by utilising the PLC relation developed by Ngeow et al. (2022) for Type II Cepheids, incorporating $[Fe/H]$ and Gaia DR3 parallax.

**Table 7.** Summary of the stellar fundamental parameters derived for DF Cygni in this study.

| Mass | Radius | Luminosity | Method |
|---|---|---|---|
| ($M_\odot$) | ($R_\odot$) | (L$_\odot$) | |
| $1.01^{+0.78}_{-0.69}$ | $33.615^{+3.899}_{-2.808}$ | – | SED |
| – | $36.53 \pm 1.39$ | $619 \pm 195$ | PR and PLC |

## 5. Discussion

DF Cyg is a bright RV Tauri-type variable star known for its challenging light curve analysis owing to its complex and ever-changing brightness. Extensive observations have been conducted for over 50 years using ground-based telescopes (AAVSO and AFOEV) and the *Kepler* spacecraft, revealing consistent short- and long-term periodic variations. The results obtained from this study are summarised in Table 7.

This study employed a different method than Bódi et al. (2016) to derive the light curves from *Kepler* data, resulting in differences showcased in Figures 7 for Q0 and 8 for Q1-Q17 data. Bódi et al. (2016) downloaded the *Kepler* dataset from The *Kepler* Asteroseismic

Science Operations(KASOC) in raw light curve mode and used an approach similar to Bányai et al. (2013) for their analysis. However, this study utilized a different method for deriving the light curves from *Kepler* TPFs (Target Pixel Files) by employing the technique described in Hartig et al. (2014). The result is consistent with the previous findings (Percy and Ursprung, 2006; Bódi et al., 2016; Vega, Stassun, Montez, Boyd and Somers, 2017; Manick et al., 2019), a long-term period of approximately 778 days was identified for DF Cyg. Accordingly, the pre-reduction and masking processes were conducted separately for each quarter of *Kepler*, with the objective of obtaining data with the least loss. The discrepancy between the light curve in this study and that obtained by Bódi et al. (2016) are shown in Figures 7 for Q0 and 8 for Q1-Q17 data, respectively. When the variable nature of the star is considered, this study achieved a reduction that minimizes the short-term effects on the pulsation period of the star. The brightness difference between derived light curves from Q0 *Kepler* data employed in this study and in Bódi et al. (2016) for DF Cyg ($\Delta m \approx 0.80$ mag) comes from the fact that Bódi et al. (2016) used the raw light curve of the star. Vega, Stassun, Montez, Boyd and Somers (2017) used the SAP data same as in this study, but they minimize systematic discontinuities in the flux due to transitions between quarters. They adjusted the data for Q0-3, Q9, and Q12. Nevertheless, adjustment of the data was not implemented in this study, as such corrections would impact the assessment of both long- and short-term changes.

Some studies have attributed long-term variations (RVb phenomenon) to a binary system (Vega, Montez, Stassun and Boyd, 2017; Bódi et al., 2016; Vega, Stassun, Montez, Boyd and Somers, 2017; Plachy et al., 2018; Manick et al., 2019); no direct evidence was found in this study. Confirmation of binarity requires long-term periodic radial velocity measurements. However, low-frequency peaks flanking the fundamental frequency (the most prominent peak in the power spectrum) are observed on both the low- and high-frequency sides. These peaks can be attributed to binarity or occultation by a circumstellar disk. Similar features were also reported in *Kepler* data by Manick et al. (2019), whose period fell within the error limits of the orbital period of a potential spectroscopic binary.

The Gaia DR3 catalogue provides another potential hint of binarity through the Renormalized Unit Weight Error (RUWE) factor. While RUWE value exceeding 1.4 is often indicative of a binary system (Belokurov et al., 2020), a study by Penoyre et al. (2022), over $\sim$22,700 binary star candidates, suggests that only 20% of binary candidates exhibit high RUWE values. Although a high RUWE for DF Cyg raises suspicion, it is important to consider that the RUWE is derived using astrometric fits (Lindegren et al., 2021). Studies by Gezer et al. (2015) and Vega, Stassun, Montez, Boyd and Somers (2017) suggested correlations between disk sources and binary in confirmed RV Tauri binaries. Similarly, this study identified a broad infrared excess in the SED of DF Cyg (Figure 6), consistent with a circumstellar disk. We estimated the disk temperature to be approximately 870 K using Wien's displacement law, which was slightly higher (i.e., 100 K) than the value reported by Vega, Stassun, Montez, Boyd and Somers (2017). Corporaal et al. (2023) demonstrated the presence of circumstellar disks around the evolved stars using mid-infrared interferometric data. However, further investigation is required to refine the disk structure of the DF Cyg.

Subtracting the long-term variations from the light curve revealed short-term variations associated with the radial pulsation and characteristic light curve structure of the RV Tauri stars. Using ground- and space-based photometric data, the period for these short-term variations was determined as 24.781 days. The double period of short-term variations was responsible for the alternating deep minimum cycle observed in the DF Cyg.

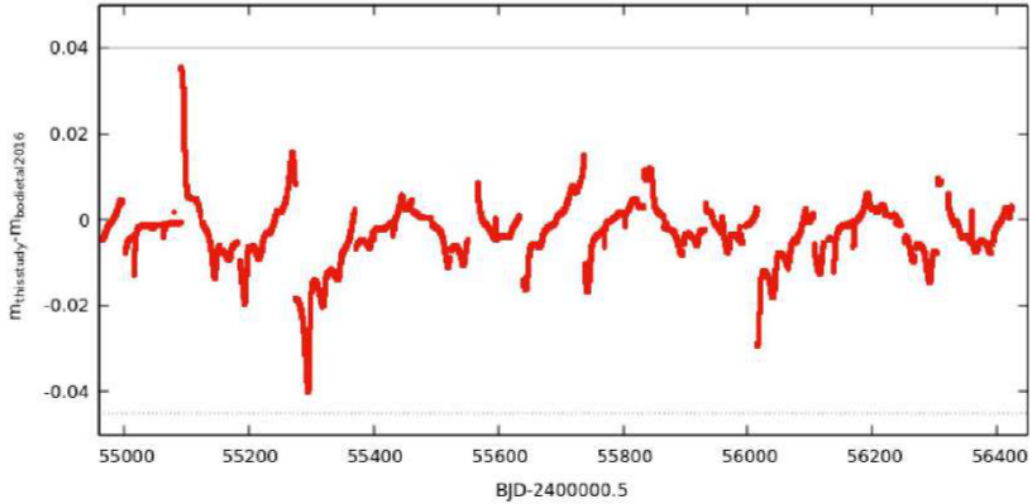

**Figure 8.** Same as Figure 7, but for Q1-Q17. The dotted and solid lines represent $\Delta m = -0.045$ mag and $\Delta m = 0.04$ mag, respectively.

By leveraging the high time resolution and precision of *TESS* (120-second cadence) data alongside *Kepler* observations, we were able to derive the subharmonics of the pulsation frequencies and uncover significant variations in pulsation amplitude. These findings are consistent with previous analyses of the *Kepler* data by Bódi et al. (2016), Kiss and Bódi (2017), and Vega, Stassun, Montez, Boyd and Somers (2017). Importantly, our *TESS* data provided new pulsation frequencies with unprecedented precision. In addition, a review of the *TESS* Data Release Notes for sectors 15, 54, and 55 revealed no indication of data anomalies, anomalous effects, or scattered light affecting the target DF Cyg (TIC 272951532).

The reported $\log g$ values for luminous RGB stars in the literature range from 0 to 2.2 cm s$^{-2}$ (Hogg et al., 2019, from APOGEE DR17). For RGB stars from LAMOST DR8, $\log g$ values are reported in the range of 0 to $0 \leq \log g \leq 0.002 T_{\rm eff} - 8.138$ cm s$^{-2}$ (Zhou et al., 2023). Based on spectroscopic and SED analyses, we determined that DF Cyg is a red giant star, which is consistent with the findings of Manick et al. (2019). However, Manick et al. (2019) fixed $\log g$ at a certain value rather than solving it. On the other hand, Mohorian et al. (2024) proposed that DF Cyg is a dusty post-RGB star that prematurely completed its low-mass RGB evolution due to binary interactions, a scenario that remains uncertain. The stellar mass calculated from the SED-ID2 ($1.01^{+0.78}_{-0.69} M_\odot$) aligns with the results of Mohorian et al. (2024). Additionally, the stellar radius derived from the PR relation and the luminosity obtained from the SED fitting method are consistent with each other and with the PL relation for the Type II Cepheid variables within the error limits. The stellar luminosity was also comparable to the value of $657 \pm 103$ L$_\odot$ reported by Mohorian et al. (2024).

Willson and Hill (1979), and Hill and Willson (1979) proposed that pulsation instability in RV Tauri stars leads to shock waves, causing line profile distortions. As the stellar atmosphere expands and contracts owing to pulsations, the line profile becomes asymmetric. Baird (1984) further explained that a shock wave occurs when the upper and lower atmospheric layers collide. This can result in absorption line splitting or doubling and emission lines from neutral helium (He I) or hydrogen-alpha (He I). While these emission lines were not detected in the McDonald spectra of DF Cyg owing to its cooler photosphere and limited wavelength coverage (4800-5600 Å), the observed chaotic pulsations and variations in the Cross-Correlation Function (CCF) and line profiles (see Figure 4) suggest the presence of shockwaves in the stellar atmosphere.

## 6. Concluding remarks

We present a comprehensive photometric, spectroscopic, and spectral energy distribution (SED) analysis of the RV Tauri star DF Cyg, combining decades of ground-based observations with new, high-precision space-based photometry from *TESS*. This multi-faceted approach allowed us to thoroughly investigate the origin of the RVb phenomenon and the star's fundamental properties. While the long-cadence *TESS* data match the precision of *Kepler*, the short-cadence *TESS* observations provide unprecedented resolution for probing short-term variations in DF Cyg. The accuracy of the Sector 14 data is nearly equal to that of the *Kepler* data, while the 120-second cadence data from sectors 15, 54, and 55 represent the most precise observations of DF Cyg to date.

By combining fifty years of ground-based data from AAVSO, AFOEV, ASAS, ASAS-SN, and SuperWASP with these new *TESS* observations, our study provides a unique and comprehensive photometric analysis across multiple wavelengths and phases. The high cadence and unprecedented precision of the *TESS* data were instrumental in allowing us to focus on the star's rapid variations. By eliminating long-term variations, we were able to clearly detect the short-term cycles of the fundamental pulsation frequency and its integer harmonics. For the first time, our analysis of the *TESS* power spectrum revealed seven new integer harmonics of the fundamental frequency (2f/9, 3f, 4f, 5f, 6f, 7f, and 8f). These new frequencies and sub-harmonics provide compelling evidence for complex internal processes, such as energy transfer between different pulsation modes, offering critical clues about the star's pulsation mechanism and internal structure. While the limited time span of the *TESS* observations restricted our ability to fully characterize long-term behavior, the data provided invaluable insights into the rapid variations and pulsational modes of DF Cyg, and also provided a more accurate fit for our SED analysis at wavelengths close to the red region. These new frequencies will thus provide observational constraints for future models of DF Cyg's internal structure and evolution. For long-term variations, we derived orbital periods of $780.1 \pm 1.1$ days (AAVSO), $773.9 \pm 0.1$ days (*Kepler*), and $779.7 \pm 1.8$ days (AFOEV) using multi-instrument data (AAVSO, AFOEV, ASAS, ASAS-SN, SuperWASP, and *Kepler*). Short-term pulsation periods were measured across datasets, yielding $24.658 \pm 0.172$ days (ASAS-N), $25.590 \pm 0.174$ days (ASAS-SN), $25.021 \pm 0.274$ days (SuperWASP), $25.084 \pm$

0.180 days (AAVSO), 25.036 ± 0.090 days (AFOEV), 24.923 ± 0.002 days (*Kepler*), and 25.282 ± 0.001 days (*TESS*).

From *TESS* sectors 14–15 and 54–55, we identified seven and eight frequencies, respectively, linked to short-term variations. Among the 33 total frequencies, 30 correspond to pulsations, while three align with the orbital cycle. Notably, seven harmonics of the fundamental frequency ($2f/9$ to $8f$) are reported here for the first time.

Our light-curve extraction from *Kepler* target pixel files (TPFs) followed the method of Hartig et al. (2014), with independent pre-reduction and masking for each quarter to minimize data loss. This approach reduced short-term artifacts in the pulsation signal compared to the raw KASOC pipeline data used by Bódi et al. (2016), as shown in Figures 7 (Q0) and 8 (Q1–Q17).

Spectroscopically, cross-correlation function (CCF) analysis revealed pulsation-phase-dependent line profile distortions, reflecting radial velocity shifts and atmospheric dynamics. We determined optimal parameters (Teff, $\log g$, v sin i, [M/H], vmic,dop, and $v_{\text{shift}}$) for individual spectra to account for short-term variability.

SED modeling uncovered a double-peaked structure, with the secondary peak indicating a circumstellar disk temperature of $\sim$ 870 K. Combining Gaia DR3 parallax with dynamic nested sampling, we derived pulsation-phase-dependent radii of $33.615^{+3.899}-2.808 R\odot$ (expanded) and $16.610^{+0.621}-0.613 R\odot$ (contracted), consistent with the pulsation radius relation ($36.53 \pm 1.39 \, \text{R}_\odot$).

Discrepancies in published parameters for DF Cygni—notably in radius (Vega, Stassun, Montez, Boyd and Somers, 2017, 10.3 ± 3.8 R⊙), luminosity (Manick et al., 2019, 990 ± 190 L⊙), and atmospheric properties (Mohorian et al., 2024)—primarily stem from methodological differences and data limitations. Our results resolve these through: (i) Gaia DR3 parallax (0.2746 ± 0.0236 mas), reducing distance errors by >50% versus earlier releases; (ii) phase-resolved SED analysis, capturing pulsation-driven radius variations (16.6–33.6 R⊙) ignored in prior static models; and (iii) updated Type II Cepheid relations (Ngeow et al. 2022; Jurkovic et al. 2023), yielding luminosity (619 ± 195 L⊙) consistent with SED-inferred values. Crucially, independent methods in this work (spectroscopy: $T_{\text{eff}} = 4781^{+66}-63$ K; SED: 5150–5360 K; PR relation: 36.5 ± 1.4 R⊙) converge within uncertainties, supporting DF Cygni's reclassification as a low-mass post-RGB system. While absolute parameters remain model-dependent, our multi-epoch approach and Gaia DR3 foundation provide a robust evolutionary framework reconciling historical inconsistencies.

Our analysis confirms that DF Cyg is a low-mass post-RGB star exhibiting RVb behavior, likely influenced by a circumstellar disk. Its atmospheric parameters resemble post-AGB stars, but its lower luminosity (200–2500 L⊙) and precise Gaia-based distance solidify its classification as a post-RGB system. This evolutionary status places DF Cyg in a unique position, bridging the gap between low-mass post-RGB systems like those in the Magellanic Clouds (Kamath et al., 2014, 2015, 2016) and higher-luminosity post-AGB stars (see Fig. D1). The first examples of post-RGB stars were reported by these authors in the Magellanic Clouds, where their luminosities ($200 - 2500$ L⊙) are significantly lower than expected for post-AGB stars, despite exhibiting similar atmospheric properties and infrared excess. These studies highlight the importance of accurate distances for calculating precise luminosities, which are well-known for the Magellanic Clouds. In this context, similar studies on sources such as DF Cyg are crucial for a better understanding of the evolution of these systems. Its evolutionary status, coupled with the presence of a circumstellar disk, has significant implications for understanding the late stages of stellar evolution and the formation of dust around evolved stars. Future observations with JWST are crucial for DF Cyg and will enable detailed characterization of the circumstellar material via high-resolution infrared data. We also anticipate that PLATO will provide long-term, high-precision photometry, ideal for investigating low-amplitude variability to better understand the photometric nature of DF Cyg. Such studies will advance our understanding of dust formation and binary evolution in low-mass post-RGB systems.

## Acknowledgments

We would like to thank Sunetra Giridhar, David L. Lambert and Enrico Corsaro for fruitful discussions and for constructive criticism of the manuscript. The Python framework (called ISPECMCMC) has been used in this study with the consent of Dr. Ahmet Dervişoğlu. The authors of the paper would like to express their gratitude to him for his contribution to the primitive form of their study. We also thank to Ferhat Güney for his contribution to spectroscopic analysis of DF Cyg with the SP_Ace code. This work was supported by the Erciyes University Scientific Research Projects Coordination Unit under grant number DOSAP MAP-2020-9749. The bibliographic service NASA Astrophysics Data System and VizieR online catalogue was used in this study. We thank the observers worldwide who contributed to the DF Cyg observations from the AAVSO and AFOEV databases and were used in this study. The *Kepler* and *TESS* mission data that are included in this study are publicly available from the Mikulski Archive for the Space Telescopes (MAST). The ASAS, AAVSO, and AFOEV photometry data were obtained from the following databases: Pigulski et al. (2009) for ASAS, https://app.aavso.org/webobs/results/?star=000-BCJ-079 for AAVSO and http://cdsarc.u-strasbg.fr/cgi-bin/afoevList?cyg/df for AFOEV, respectively. Furthermore, the processed photometric and spectral data underlying the corresponding author will share this paper upon reasonable request.

## Appendix A. MCMC Results of Spectroscopic Analysis of DF Cyg

## Appendix B. ARIADNE SED fitting results of DF Cyg

## Appendix C. McDonald spectra of DF Cyg

## Appendix D. Evolutionary HR Diagram of post-RGB and post-AGB stars

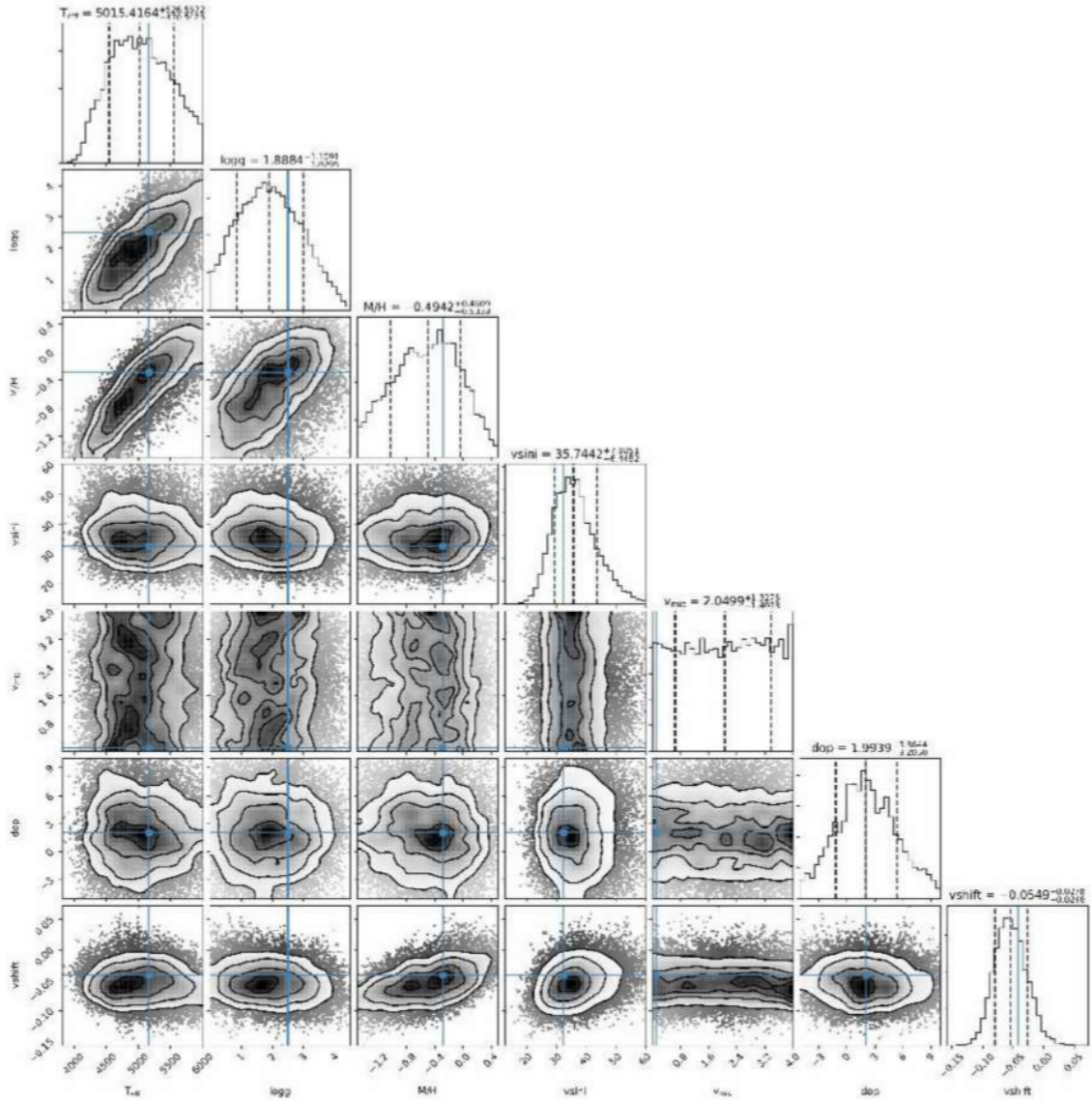

**Figure A1.** MCMC results of the spectral parameters and accompanying uncertainties for observed spectrum on 16th November 2008 of DF Cyg. $T_{eff}$, $\log g$, [M/H], and vsini were strongly correlated. Solid contours represent posterior distribution densities. The histogram distributions represented by solid lines are plotted across the associated quantities with the 16th, 50th, and 84th percentile levels with dashed lines. $T_{eff}$ is given in K units, $\log g$ and [M/H] are in dex units, and vsini, $v_{mic}$, dop, and $v_{shift}$ are in km s$^{-1}$ units.

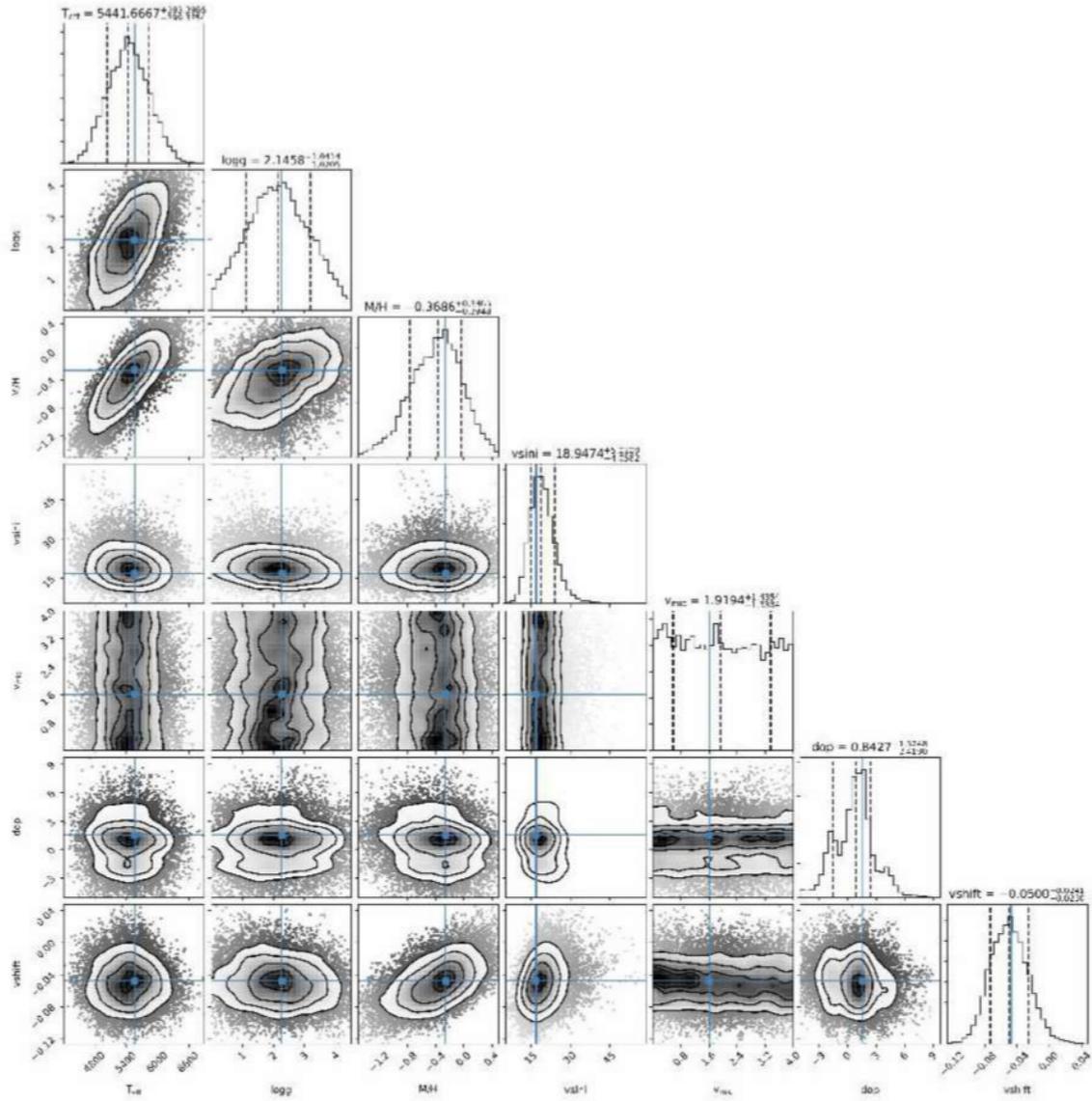

**Figure A2.** Same as Figure A1, but for spectroscopic observation of DF Cyg on September 16, 2008.

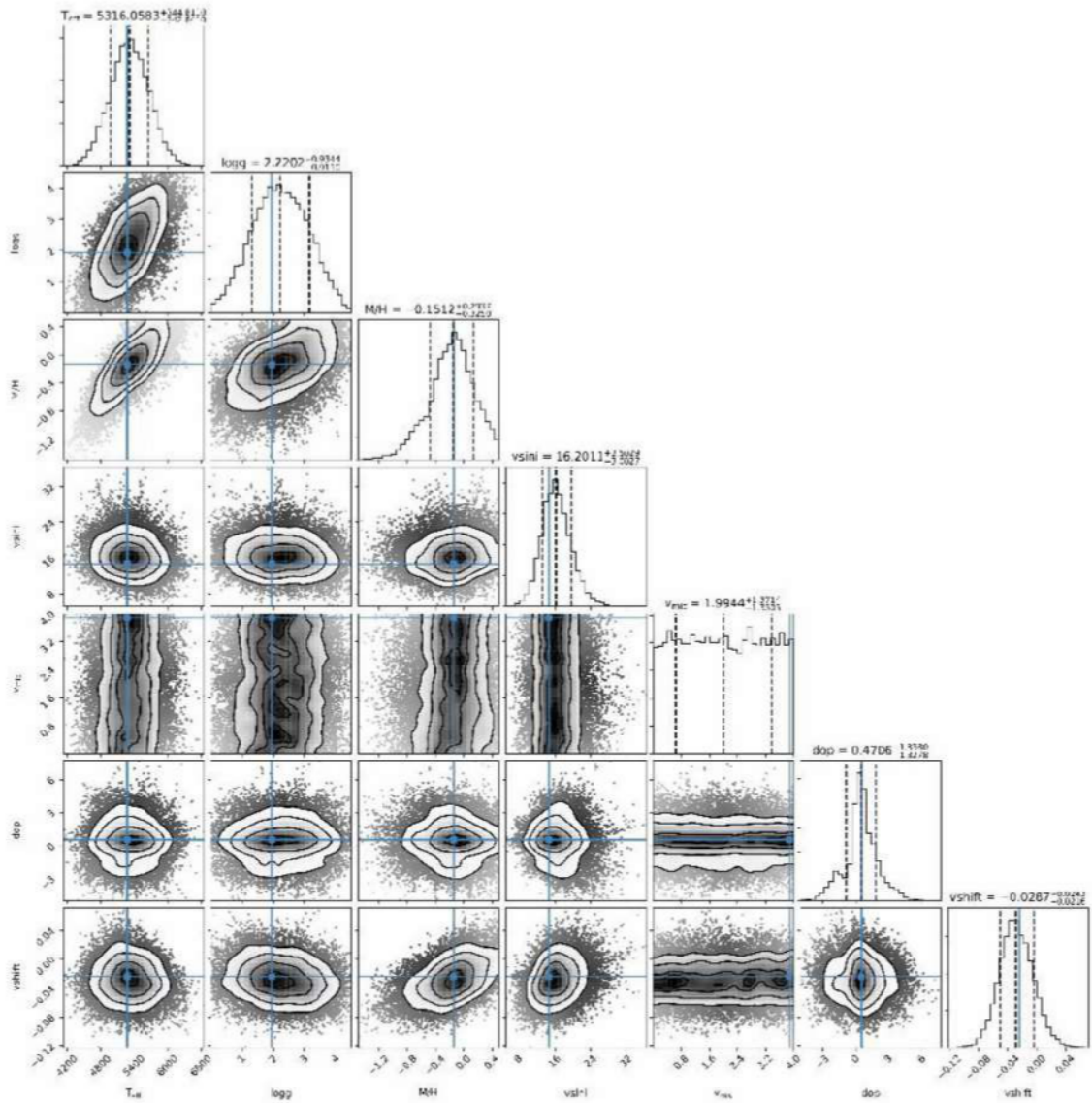

**Figure A3.** Same as Figure A1, but for spectroscopic observation of DF Cyg on September 18, 2008.

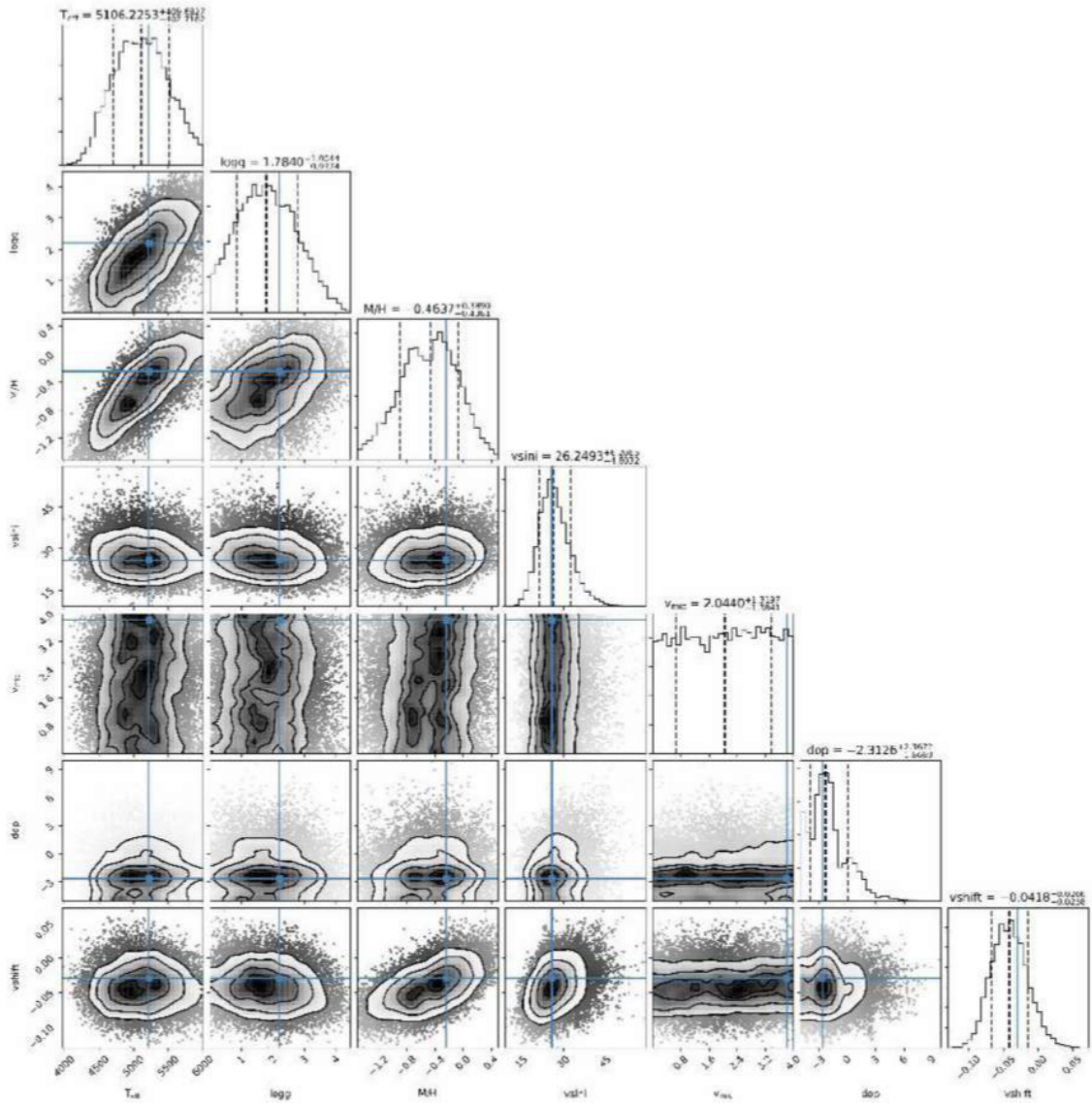

**Figure A4.** Same as Figure A1, but for spectroscopic observation of DF Cyg on October 19, 2008.

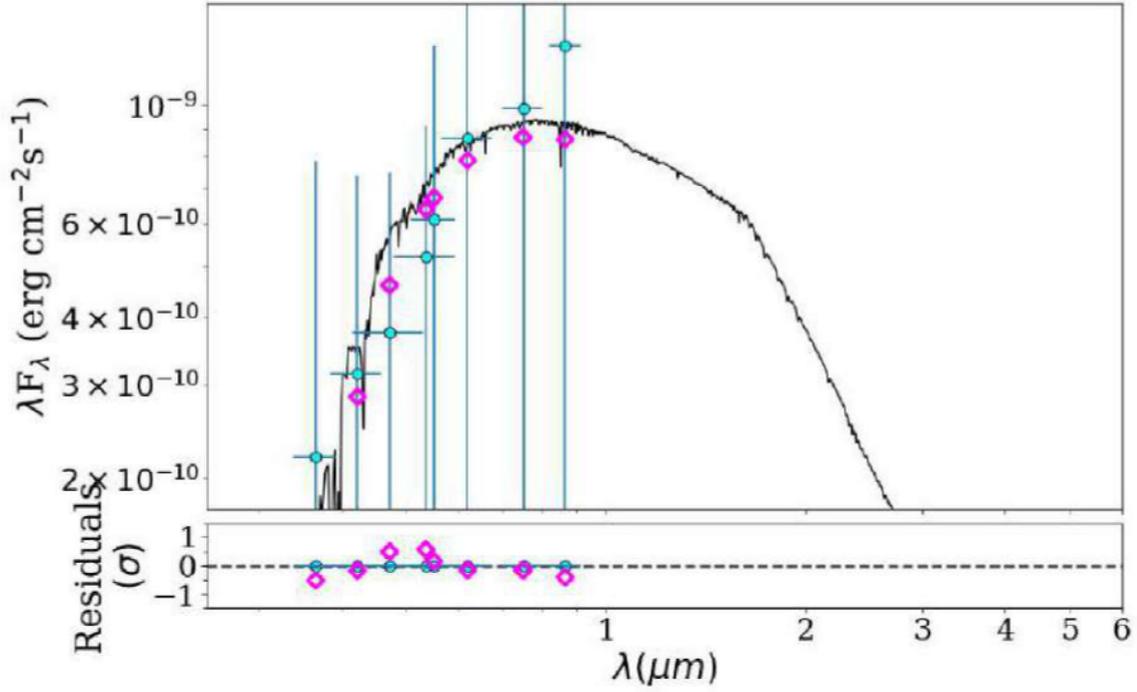

**Figure B1.** *Top:* ARIADNE reddened model of SED-ID1 and photometric observation of DF Cyg. SED-ID1 includes Johnson, Tycho, and SDSS photometric observations, and their error bars of the star (blue circles) are at the increased to the maximum brightness level of the long-term variations in the star. The diamonds denote synthetic photometry. *Bottom:* Residuals of the fit that are scaled to normalize the photometry errors.

**Table D1.** Various neutral and ionised iron lines showing clear signs of shock-induced behaviour in McDonald spectra of DF Cyg.

| Species | $\lambda$ (Å) | Species | $\lambda$ (Å) |
|---|---|---|---|
| Fe I | 5001.87 | Fe I | 5307.37 |
| Fe I | 5002.80 | Fe I | 5322.05 |
| Fe I | 5014.95 | Fe I | 5307.37 |
| Fe I | 5022.24 | Fe I | 5322.05 |
| Fe I | 5121.65 | Fe I | 5339.94 |
| Fe I | 5141.75 | Fe I | 5367.48 |
| Fe I | 5145.10 | Fe I | 5369.97 |
| Fe I | 5150.85 | Fe I | 5373.71 |
| Fe I | 5191.47 | Fe I | 5379.58 |
| Fe I | 5202.35 | Fe I | 5383.38 |
| Fe I | 5215.19 | Fe I | 5415.21 |
| Fe I | 5217.40 | Fe II | 5132.67 |
| Fe I | 5302.31 | Fe II | 5325.56 |

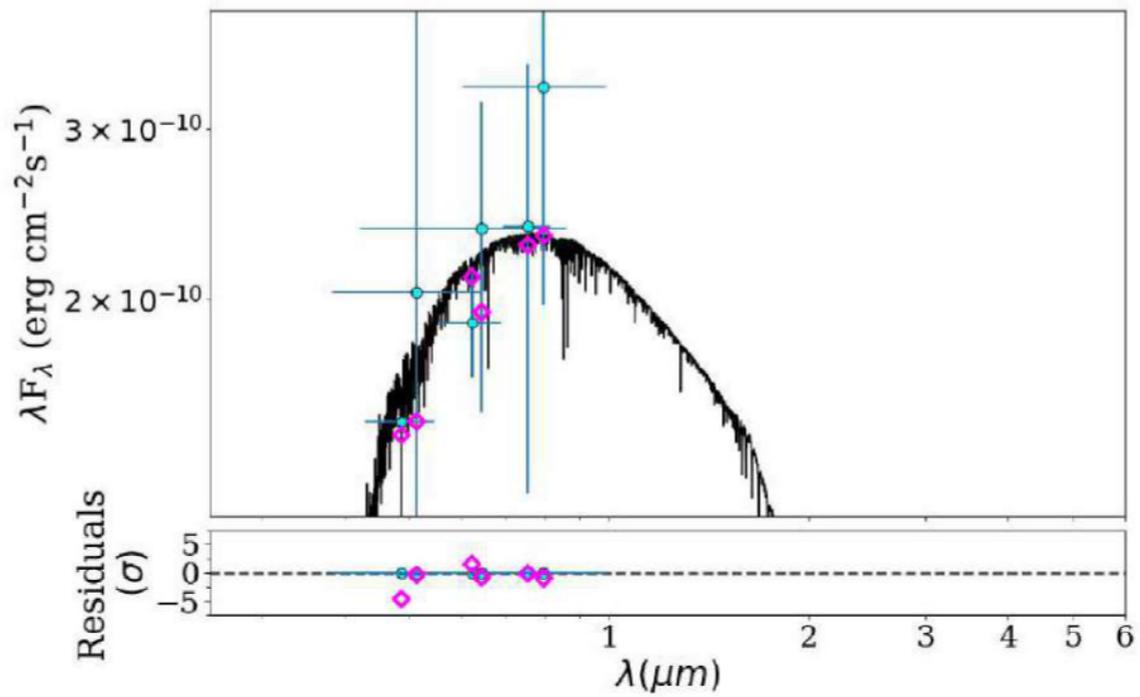

**Figure B2.** Same as Figure B1, but for SED-ID2 which includes *Gaia DR3*, *PS1*, and *TESS* photometric observations and their error bars of the star (blue circles) is at the decreasing to the lowest level of the long-term variations of the star.

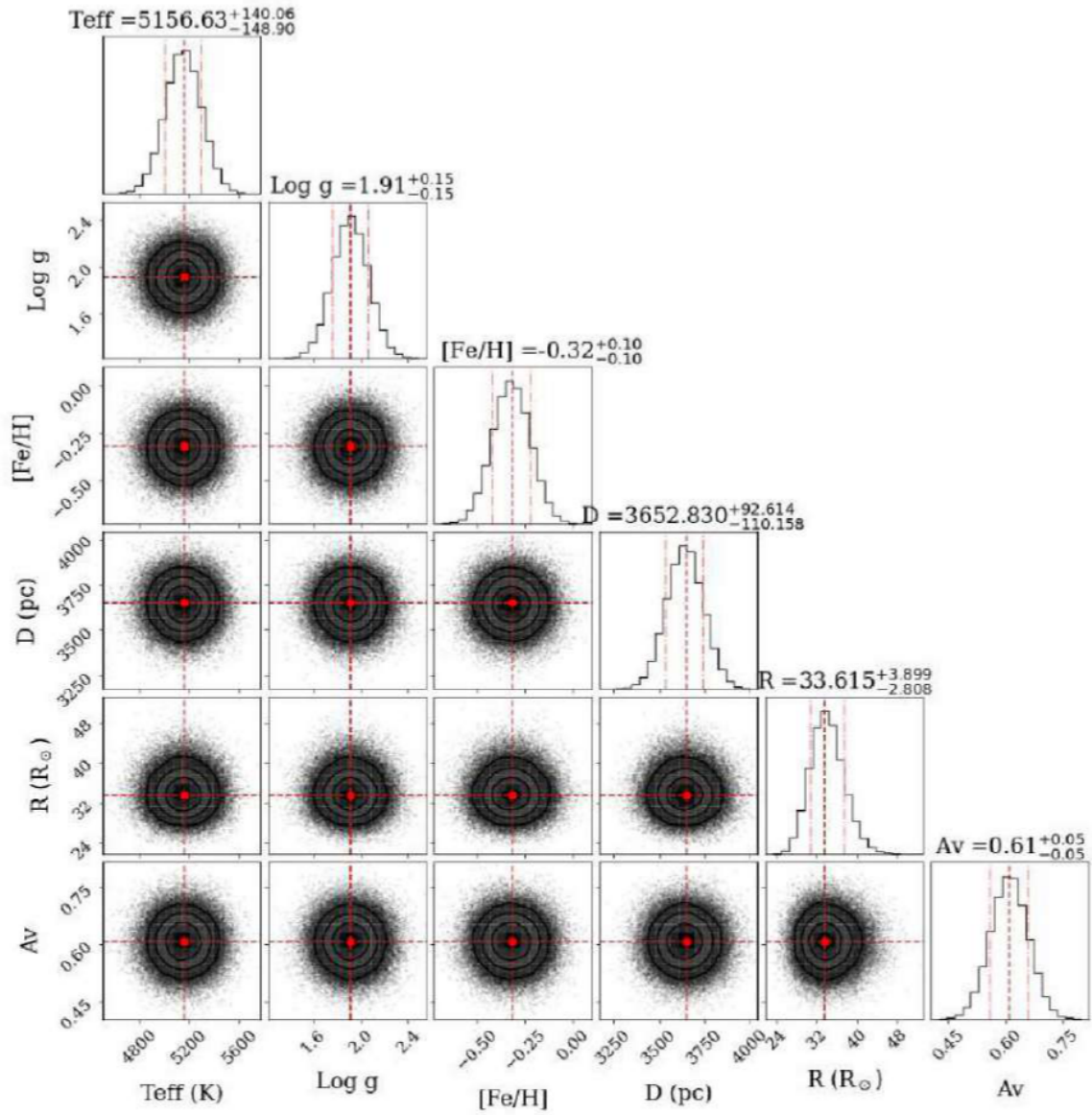

**Figure B3.** ARIADNE SED fitting results of DF Cyg. The parameters and accompanying uncertainties in the SED model of the SED-ID1 observed photometric points. Effective temperature (Teff) in the K unit, logarithmic surface gravity ($\log g$) in the dex unit, metallicity ([Fe/H]) in the dex unit, distance (D) in the pc unit, radius (R) in the solar unit and interstellar extinction (Av) in unitless of SED-ID1 model results of DF Cyg are plotted.

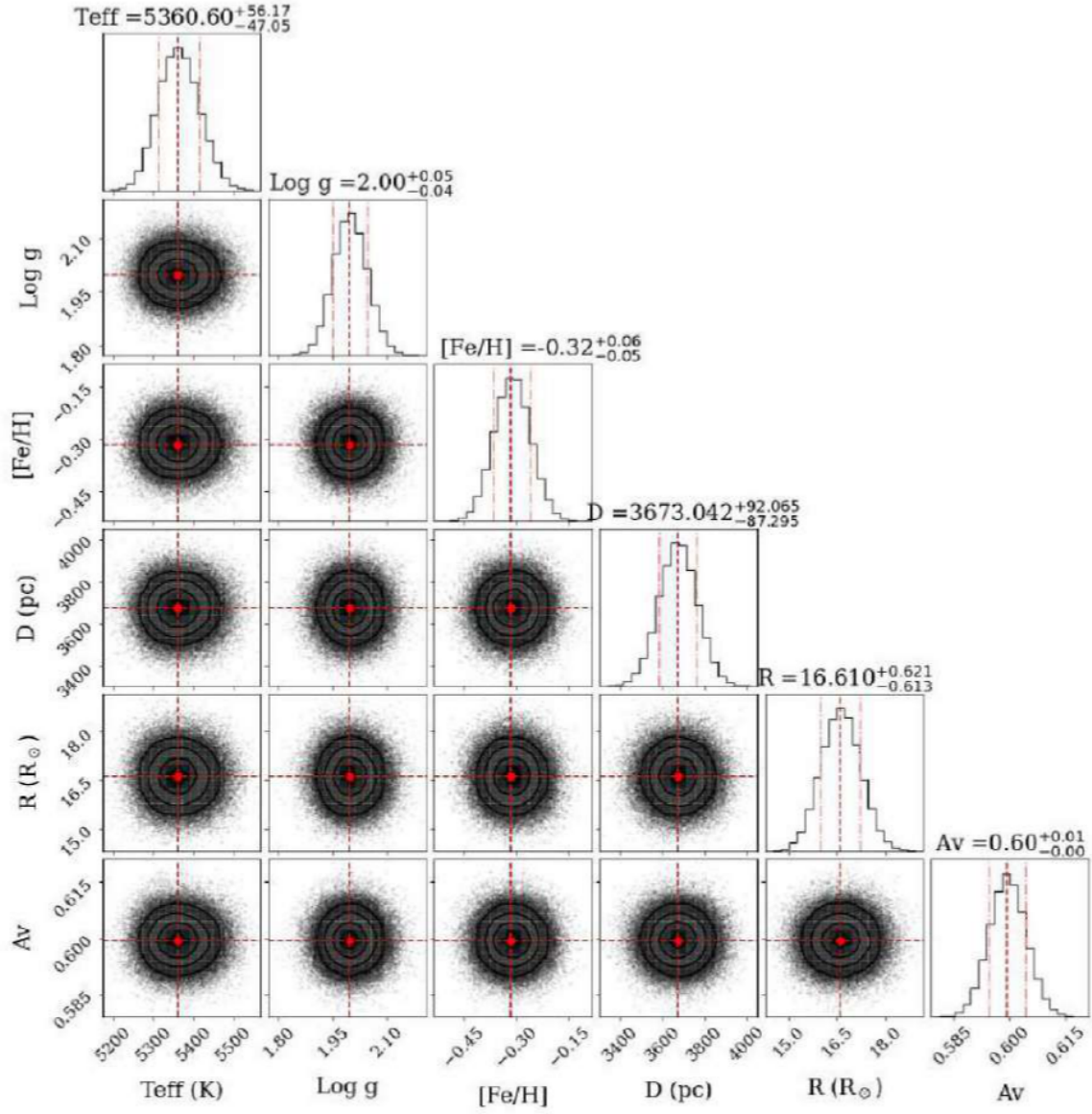

**Figure B4.** Same as Figure B3, but for SED-ID2.

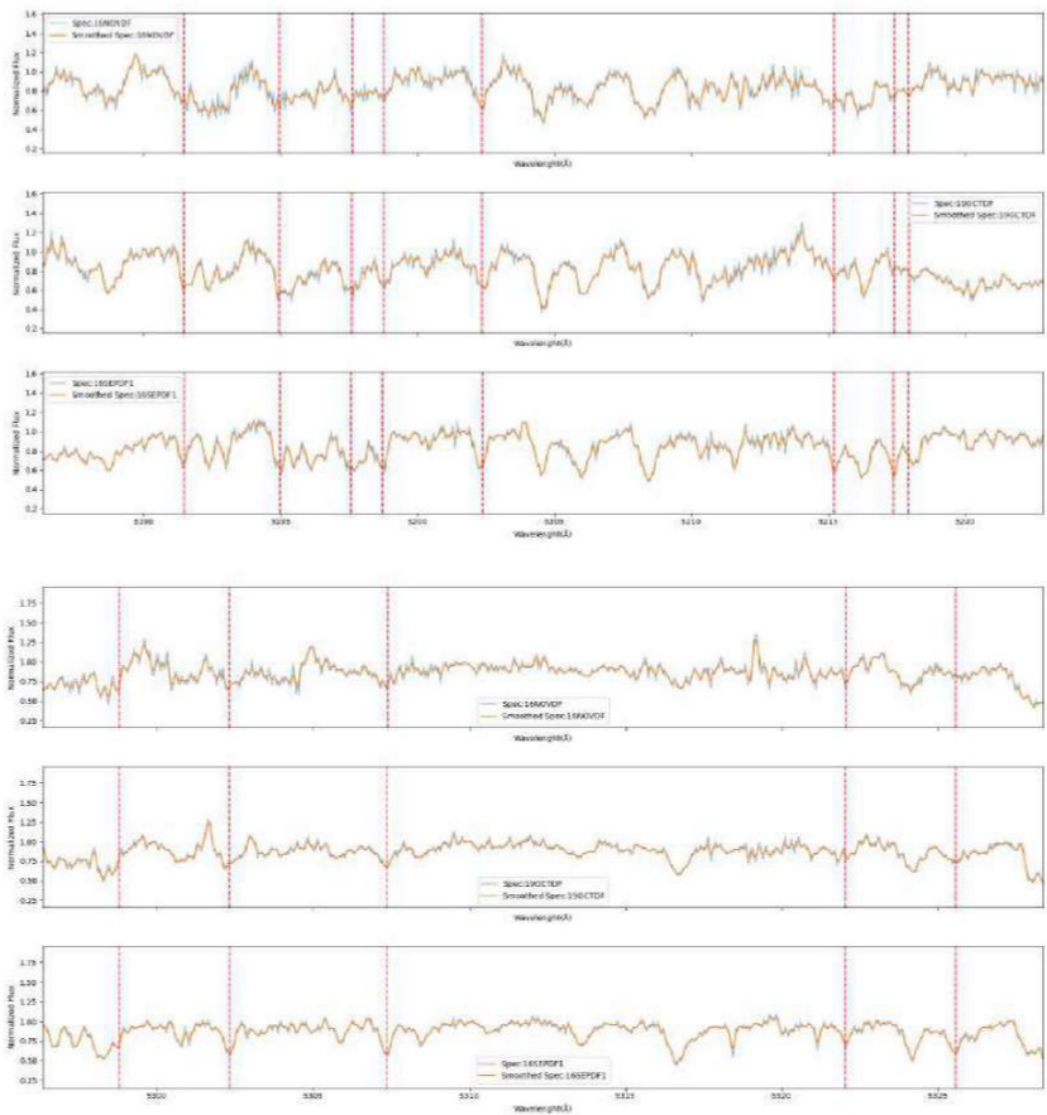

**Figure C1.** The McDonald spectra of DF Cyg obtained on HJD 2454726.66991635 (September 16, 2008), HJD 2454759.64179477 (October 19, 2008), and HJD 2454787.65328437 (November 16, 2008) at several separate wavelength regions. Neutral and ionized Fe lines are indicated.

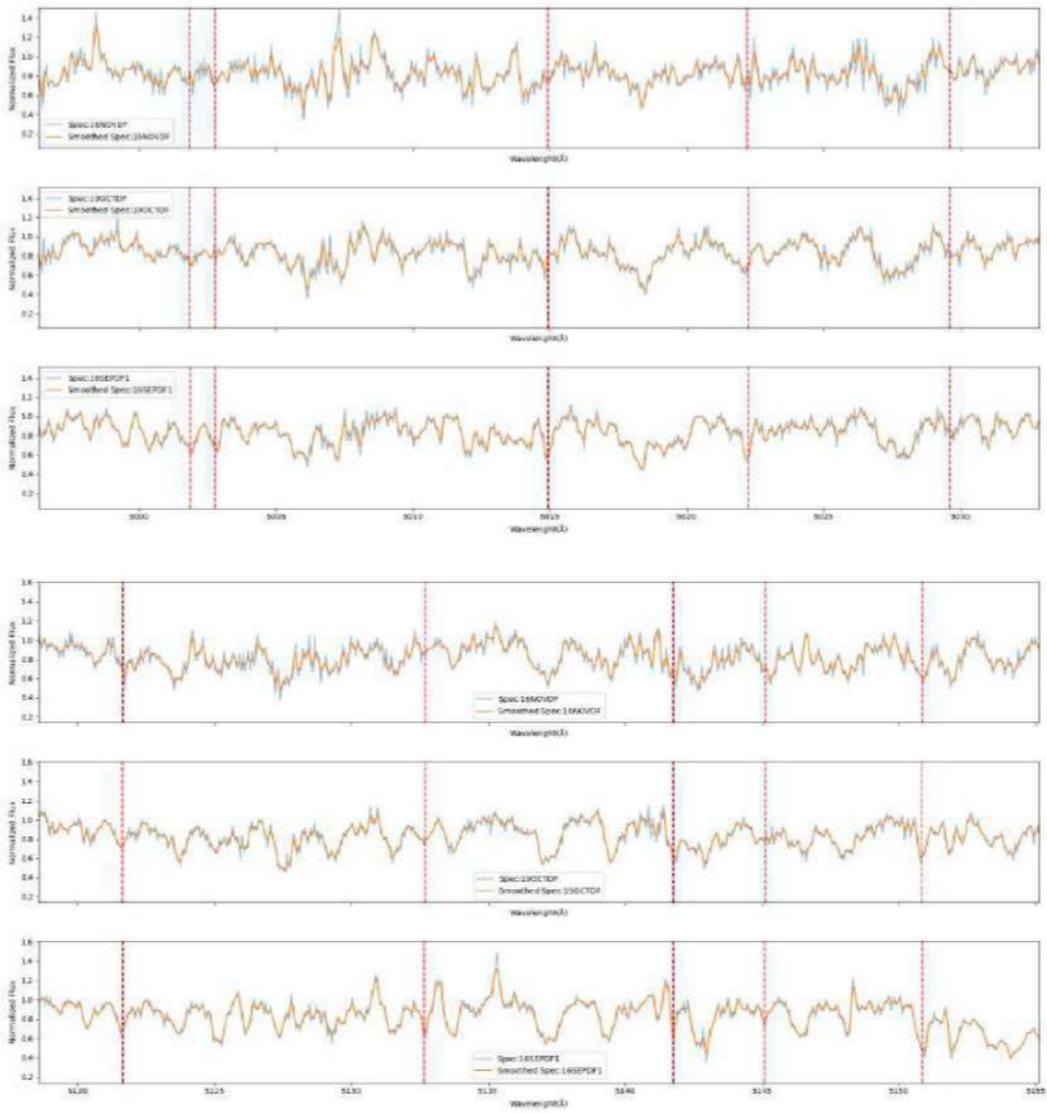

**Figure C2.** Continuation of Figure C1.

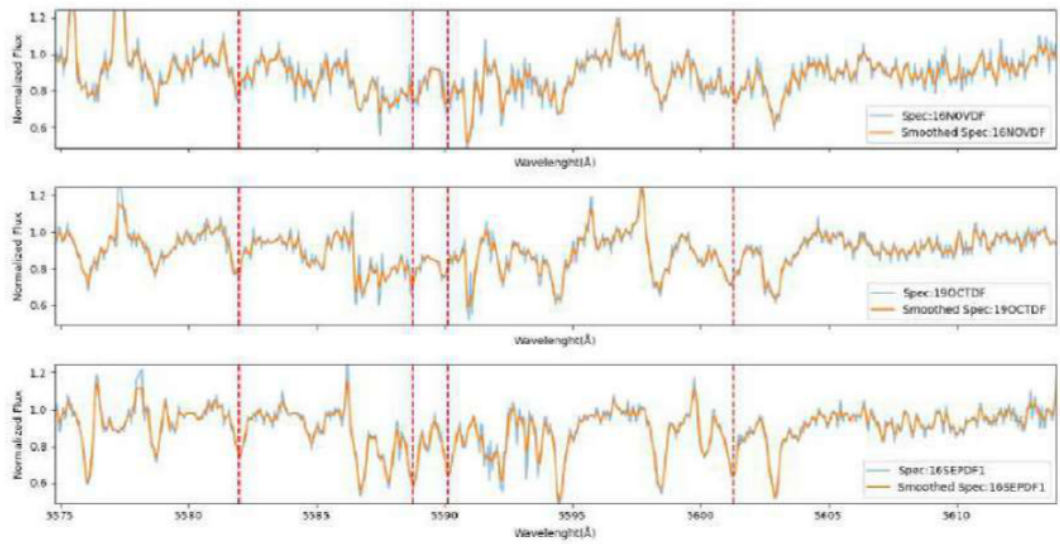

**Figure C3.** The McDonald spectra of DF Cyg obtained on HJD 2454726.66991635 (September 16, 2008), HJD 2454759.64179477 (October 19, 2008), and HJD 2454787.65328437 (November 16, 2008) at several separate wavelength regions. Ca lines are indicated.

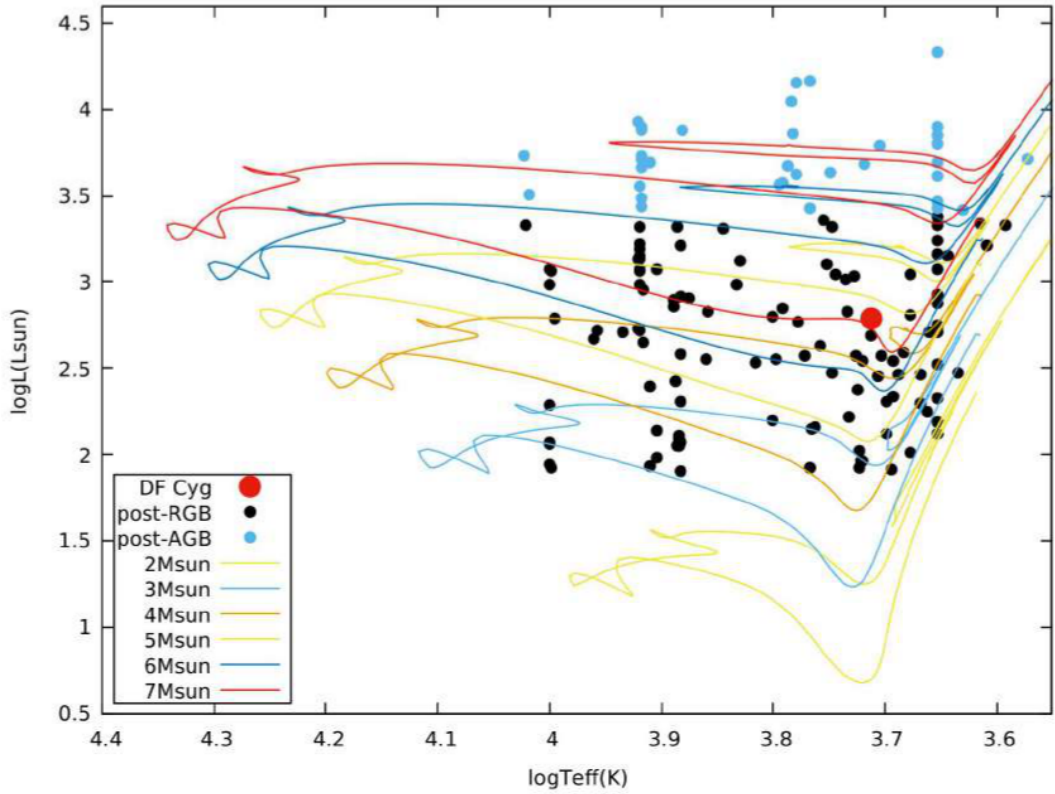

**Figure D1.** Evolutionary HR diagram of post-RGB (black filled circle) and post-AGB (blue filled circle) stars. The data of the post-RGB and post-AGB stars are taken from Kamath et al. (2015). The stellar evolutionary models are generated with Modules for Experiments in Stellar Astrophysics (MESA) Isochrones and Stellar Tracks (MIST) project (Choi et al., 2016). The models consists of masses ranging from 2 to $7 M_\odot$. The red filled circle denotes DF Cyg.